\documentclass[twocolumn,aps,prb,superscriptaddress,nofootinbib]{revtex4-2}
\usepackage{chngcntr}
\usepackage{graphicx}
\usepackage{color}
\usepackage{amsmath}
\usepackage{enumitem}
\usepackage{amssymb}
\usepackage{cancel}
\usepackage{ulem}
\usepackage{multirow}
\usepackage[colorlinks=true,linkcolor=blue,citecolor=blue]{hyperref}
\newcommand{\be}{\begin{equation}}
\newcommand{\ee}{\end{equation}}

\newcommand{\bea}{\begin{eqnarray}}
\newcommand{\eea}{\end{eqnarray}}

\usepackage[dvipsnames]{xcolor}

\definecolor{revorange}{RGB}{224,138,0}
\definecolor{revgreen}{RGB}{0,140,60}

\definecolor{revteal}{RGB}{0,130,130}

\definecolor{revpurple}{RGB}{125,40,175}
\definecolor{revbrown}{RGB}{150,75,0}
\definecolor{revolive}{RGB}{105,118,0}
\definecolor{revsky}{RGB}{0,140,210}
\definecolor{revpink}{RGB}{224,69,123}
\usepackage{pifont}
\newcommand{\cnum}[1]{\ding{\numexpr171+#1\relax}}%

\begin{document}

\title{
Symmetry-Enabled Rolling of Composite Bosons: A Route to Superconductivity from Strong Repulsion with $T_c$ Proportional to the Electron Bandwidth
}
\author{Zhiyu Dong}
\affiliation{Department of Physics and Institute for Quantum Information and Matter, California Institute of Technology, Pasadena, California 91125}

\begin{abstract}
	Motivated by superconductors discovered in systems with a strong Coulomb repulsion, we ask whether such a repulsion can give rise to superconductivity with a transition temperature $T_c$ bounded only by the electron bandwidth. Coulomb repulsion can produce pairing 
	as it binds electrons to holes. If each hole binds three electrons into a three-electron-one-hole (3e1h) composite, the condensation of these bosonic composites can give rise to superconductivity. However, the resulting $T_c$ falls far below the electron bandwidth in the controlled regime where the binding is much stronger than the carriers' kinetic energy. This is because, on a lattice, a tightly bound composite usually has to move through a higher-order process where the intermediate state costs an energy of order the binding energy, so that the strong binding makes the composite heavy, limiting the superfluid stiffness and hence $T_c$. Here, we show that symmetry offers a way around this tradeoff. In lattices satisfying certain symmetry requirements, a composite can move by single-particle hops without any costly intermediate state, a motion we call \emph{rolling}. Such rolling keeps the mass of the composite finite even for an infinitely strong binding, so that a $T_c$ proportional to the electron bandwidth can be achieved in the controlled regime. We derive a criterion for rolling and apply it to construct all fourteen minimal lattice models that support such rolling. 
We show that rollable 3e1h composites can indeed be stabilized in the ground state.
We then propose to realize such superconductivity in a dual-gate bilayer,
and use exact numerics to show that $T_c$ can reach a few percent of the electron hopping.
These results establish rollability as a guiding principle in the search for and design of superconductors in strongly repelling electron systems, with the minimal models constructed here as prototypes.
\end{abstract}
\maketitle

\tableofcontents

\section{Introduction}
\label{sec:introduction}

What limits the superconducting transition temperature $T_c$ is a question of fundamental importance.
In the theory of Bardeen, Cooper, and Schrieffer~\cite{bcs_1957} and in its extension to strong coupling~\cite{mcmillan_1968,allen_dynes_1975,cohen_anderson_1972,esterlis_2018_breakdown,esterlis_kivelson_scalapino_2018,chubukov_2020,hofmann_2022}, $T_c$ is limited by the bandwidth of the pairing interaction, which is the Debye frequency for phonon-mediated pairing. A pairing interaction with a broader bandwidth would therefore allow a higher $T_c$. The Coulomb interaction between electrons is promising in this respect if it can mediate pairing, because it acts over the entire electron bandwidth. 
But the Coulomb interaction is repulsive, so how can it be turned into a pairing glue? People have proposed many theories aiming at this goal%
~\cite{monthoux_pines_lonzarich_2007,scalapino_2012,maiti_chubukov_2013}.
The classic example is the Kohn--Luttinger mechanism~\cite{kohn_luttinger_1965}, in which the screened Coulomb interaction gives an attraction at second order, which exceeds the first-order repulsion in pairing channels of large angular momentum $\ell$. However, this attraction decays quickly as $\ell^{-4}$. Later work found that the attraction survives down to $\ell=1$ in dilute gases, where diluteness takes the place of large $\ell$ in making the first-order repulsion weaker than the second-order attraction~\cite{fay_layzer_1968,kagan_chubukov_1988,kagan_1991,baranov_efremov_kagan_1993,kagan_2016}.
The mechanism has also been extended to two-dimensional systems~\cite{chubukov_1993,chubukov_kivelson_2017} and to lattice models~\cite{raghu_kivelson_scalapino_2010,kagan_2014}, and studied with renormalization-group methods~\cite{zanchi_schulz_1996,zanchi_schulz_2000,nandkishore_levitov_chubukov_2012,maiti_chubukov_2013,fujimoto_2025,keles_li_zhao_2024}. In these theories, a $T_c$ set by the electronic energy scale can be reached in the marginally controlled regime~\cite{kagan_2014,chubukov_kivelson_2017}. There are other attempts aiming to capture instabilities and collective modes, such as spin fluctuations~\cite{anderson_brinkman_1973,scalapino_loh_hirsch_1986,miyake_schmitt_rink_varma_1986,beal_monod_bourbonnais_emery_1986,monthoux_pines_lonzarich_2007,scalapino_2012} and plasmons~\cite{takada_1978,rietschel_sham_1983}, usually either by writing down the interaction they mediate phenomenologically, or by summing an infinite series of diagrams in the random phase approximation (RPA). Recently, RPA has also been applied to flavor-polarized flat-band systems, where the interaction is very strong~\cite{chou_zhu_dassarma_2025,geier_davydova_fu_2024,yang_zhang_2025,qin_wu_2026,jahin_lin_2024,parra_martinez_2025,guerci_abouelkomsan_fu_2025}. But, strictly speaking, there is no small parameter to justify keeping only the RPA diagrams~\cite{maiti_chubukov_2013,jahin_lin_2024}, so the attraction obtained this way is not reliable. Yet, in such flavor-polarized systems, it has been shown~\cite{dong_lee_2026,dong_lee_2025} that the perturbation theory can be controlled by another small parameter even when the bare interaction is strong. In some settings, the first-order repulsion exactly vanishes whereas the second-order diagrams produce an attraction, leading to a $T_c$ set by the electronic energy scale in the marginally controlled regime.
There are also other routes, such as the exchange interaction in doped Mott insulators~\cite{anderson_1987,dagotto_1994,lee_nagaosa_wen_2006}.

Recently, superconductivity has been discovered in many 2D flat-band systems~\cite{balents_2020}, such as moir\'e graphene~\cite{cao_2018_sc}, moir\'e transition-metal dichalcogenides~\cite{xia_2025_twse2,guo_2025_twse2}, and rhombohedral graphene~\cite{zhou_2021_rtg,zhou_2021_rtg_sc}. Notably, in moir\'e transition-metal dichalcogenides and in rhombohedral graphene, electrons can even form Wigner crystals~\cite{regan2020mott,xu2020correlated,li2021imaging,han_2026_wigner}, which indicates an extremely strong Coulomb interaction~\cite{tanatar_ceperley_1989,drummond_needs_2009}. The $T_c$ in these platforms is at most a few kelvin. While this is not nominally high, it is already appreciable compared to the kinetic energy of the carriers in these systems, namely, $10^{-3}$ to $10^{-2}$ of the carriers' bandwidth.
These observations motivate us to ask two questions. 
Can an electron-electron repulsion that is much stronger than the electrons' kinetic energy be turned into a pairing interaction? If so, can it achieve a $T_c$ bounded only by the electron bandwidth?

For the first question, the answer is yes: there is indeed a scenario where a strong repulsion can be directly turned into an attraction in real space.
The basic idea is that a hole attracts electrons through the Coulomb interaction and can thereby mediate an attraction between them.
Concretely, in a half-filled two-band system, a hole in the filled lower band can bind to itself electrons in the empty upper band. If each hole binds three electrons, resulting in a three-electron-one-hole (3e1h) composite particle with charge $2e$ and bosonic statistics, then a superfluid of such composite bosons can give rise to superconductivity. Such a theory is controlled when the interaction overwhelms the electrons' kinetic energy, which is the regime of interest. This scenario has been studied and developed recently for moir\'e materials and other systems~\cite{slagle_fu_2020,crepel_fu_2021,crepel_fu_2022,crepel_2022,guerci_fu_2025}. It can also be viewed as an exciton mediating the attraction between two electrons~\cite{slagle_fu_2020,crepel_fu_2022,guerci_fu_2025}, an idea that can be traced back to the proposals of Little and of Ginzburg~\cite{little_1964,ginzburg_1964} and was later developed for metal-semiconductor interfaces~\cite{allender_bray_bardeen_1973} and for the cuprates~\cite{varma_schmitt_rink_abrahams_1987,hirsch_tang_loh_scalapino_1988}. 

\begin{figure}[!tb]
\centering
\includegraphics[width=\columnwidth]{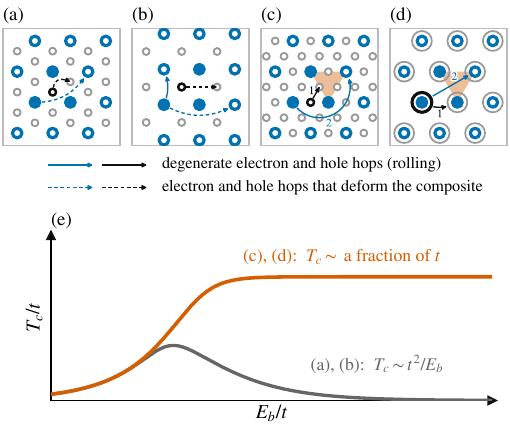}
\caption{Rollability of 3e1h composites results in distinct behaviors of superconducting $T_c$. Filled blue circles and the black ring are the electrons and the hole, and open blue and gray circles are empty electron and hole orbitals. Generic composites, such as the examples shown in (a) and (b), are ``not rollable'': To move through the plane, they must go through a two-step process (dashed arrows) with a deformed intermediate state, which costs an energy of order the binding energy $E_b$. The degenerate hop in (b) (solid arrow) does not help because it only moves the composite back and forth. Panels (c) and (d) show two examples where composites are rollable as they can move through the plane by degenerate hops without deformation.
	In (c), the electrons live on a triangular lattice, whereas the hole orbitals are located at the bond midpoints of this lattice, forming a kagome lattice. 
	In (d), each hole orbital sits on top of an electron orbital and is drawn as the larger gray ring around it. In both (c) and (d), the composite can move onto a neighboring plaquette (the orange triangle) via two successive degenerate hops, in the order indicated by the numbers.
	(e)~Distinct dependence of $T_c$ on binding strength $E_b$ for superconductivity achieved by composites that are rollable and not rollable. Upon increasing binding, the $T_c$ of the former increases until it saturates at a fraction of the hopping $t$, whereas that of the latter is suppressed.}

\label{fig:intro_rolling}
\end{figure}

For the second question, however, as already noticed in this line of work~\cite{slagle_fu_2020,guerci_fu_2025}, the $T_c$ achieved this way in general falls far below the electron bandwidth in the strong-coupling regime, where the theory is controlled. In short, this is due to a tradeoff between a composite's binding and its mass. To illustrate this point, we consider how a tightly bound 3e1h composite such as the one in Fig.~\ref{fig:intro_rolling}(a) moves on a lattice. Throughout this paper, we focus on the general case where the electron and hole orbitals are well localized, so that projecting the Coulomb interaction onto these orbitals yields a negligible pair hopping term, in which the interaction moves two particles at once. 	
The tradeoff we will illustrate below does not directly apply to cases with nonnegligible pair hopping, such as bipolaron systems with a Peierls electron-phonon coupling, which modulates the hopping itself~\cite{sous_2018,zhang_2023}, and flat bands with a nontrivial quantum geometry, whose orbitals cannot be well localized~\cite{peotta_torma_2015,tovmasyan_peotta_torma_huber_2016,torma_liang_peotta_2018}.
Thus, in the systems of our interest, a composite has to move by single hops of electrons and holes, one particle at a time.
No matter whether an electron or the hole hops first (dashed arrows in Fig.~\ref{fig:intro_rolling}(a)), the composite is deformed, which costs an energy of order the binding energy $E_b$. A second hop then restores its shape at a new position. This composite thus can only move through a two-step process with a deformed intermediate state, so its effective hopping is only of order $t^2/E_b$, where we use $t$ to denote the electron and hole hopping amplitudes within this introduction for simplicity. A generic composite (e.g., the one in Fig.~\ref{fig:intro_rolling}(b)) behaves in the same way, as the deformation usually occurs in an intermediate state. 
	Therefore, usually, the stronger the binding, the heavier the composites are. Their superfluid stiffness, which sets $T_c$~\cite{nelson_kosterlitz_1977,emery_kivelson_1995,uemura_1989}, drops accordingly as it is inversely proportional to the mass, so $T_c$ is suppressed at strong binding [gray curve in Fig.~\ref{fig:intro_rolling}(e)]. 
	Similar problems occur for other types of composites, such as tightly bound pairs in the attractive Hubbard model~\cite{nozieres_schmitt_rink_1985,micnas_ranninger_robaszkiewicz_1990,keller_metzner_schollwock_2001,chen_2024_bcs_bec} and small bipolarons~\cite{chakraverty_1998,esterlis_2018_breakdown,esterlis_kivelson_scalapino_2018}, which limits $T_c$ in general lattices. We note parenthetically that this binding-versus-mass tradeoff occurs only on a lattice, as in free space the mass of a composite is locked to the total mass of all its particles by Galilean invariance.

However, we find that there is a way around this tradeoff on a lattice.
Specifically, we will show that in lattices satisfying certain symmetry requirements, a composite can move by single-particle hops that never climb an interaction-scale barrier, so that 
its mass does not grow with the binding energy.
	To see that this is possible, we show in Fig.~\ref{fig:intro_rolling}(c) and (d) two lattice models of this kind, along with the pattern of the 3e1h composite on each. Here, the %
	composite can move onto a neighboring plaquette (the orange triangle) via two successive hops (black and blue arrows), in the order indicated by the labels.
	 Both are hops in the model that take the composite to an exactly degenerate state, called degenerate hops. They are degenerate as the composite is not deformed (it changes neither its interparticle distances nor how it sits on the lattice). Note that the electron hop~2 is not degenerate for the drawn composite pattern, but is degenerate for the pattern after the hole hop~1.
In both cases, the composite can be carried through the whole plane by a sequence of such degenerate hops. %
We call a composite's motion through these shape-preserving degenerate hops \emph{rolling},
in analogy to the rolling of a classical rigid body.
With such ``2D rollability'', the composite has an effective hopping of the order of the single-particle hoppings. As a result, $T_c$ can monotonically approach a fraction of the electron bandwidth, even deep in the strong-coupling regime, where the theory is controlled (see the orange curve in Fig.~\ref{fig:intro_rolling}(e)).
These examples demonstrate a route to search for natural or synthetic superconducting platforms %
in which the Coulomb interaction can reliably achieve a $T_c$ bounded only by the electron bandwidth. This motivates us to look for all settings that host a 2D-rollable 3e1h composite.

It is noteworthy that an example of rolling motion has been found and studied in the context of bipolarons, where the composite boson is formed by two electrons sharing a lattice distortion that spreads over several sites, so that the pair can move by so-called crab motion, i.e., one electron pivoting around the other%
~\cite{alexandrov_kornilovitch_2002,hague_2007,hague_etal_2007_jpcm}. The same motion has also been found for pairs bound by an intersite attraction in extended Hubbard models~\cite{adebanjo_kornilovitch_hague_2022,adebanjo_hague_kornilovitch_2024}. However, to the best of our knowledge, 
the essence of rolling motion has never been revealed, and a theory that can guide the search for platforms that support rolling composites has not been established.
For our purpose, such a theory is indispensable, as most 3e1h patterns cannot roll, and there are too many degrees of freedom to search, such as the lattice type, the shape of the composite, and the orbitals of the electrons and the hole, so one has to search systematically rather than by random enumeration.

\begin{figure}[!tb]
\centering
\includegraphics{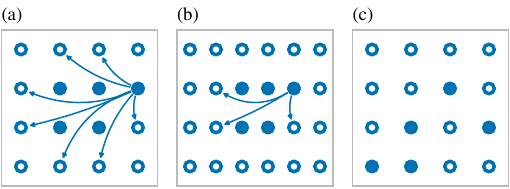}
\caption{Degenerate hops are set by the symmetry of the lattice and the shape of the composite. Blue dots and circles represent occupied and empty electron orbitals. (a)~On a square lattice, one electron of this five-electron composite has seven degenerate hops (arrows), each taking the composite to its image under a reflection or a rotation of the lattice. (b)~On a rectangular lattice, which has fewer symmetries, three degenerate hops remain. (c)~A generic composite can have no degenerate hop if its pattern is too irregular.}

\label{fig:intro_hop_symmetry}
\end{figure}

\begin{figure}[!tb]
\centering
\includegraphics[width=\columnwidth]{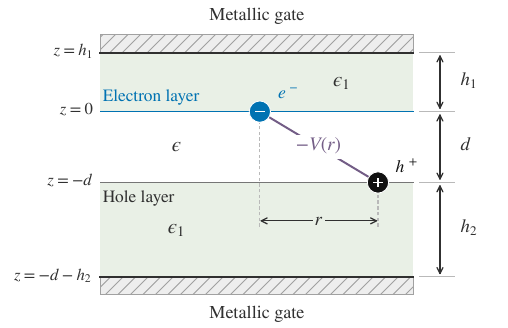}
\caption{Realization in dual-gate bilayer geometry. The top and bottom gates lie at distances $h_1$ and $h_2$ from the sample, respectively. In the sample, there are electron orbitals and hole orbitals sitting in two layers a distance $d$ apart.
The space between each gate and the sample is filled with a dielectric spacer of permittivity $\epsilon_1$, and the space between the layers has the permittivity $\epsilon$. %
}	

\label{fig:single_gate_geometry}
\end{figure}

In this paper, we will work out such a theory of rolling, and use it to construct systems with rollable 3e1h composites.
We will start by showing that rollability is essentially tied to symmetry. This is intuitively reasonable, as rollability roughly depends on which degenerate hops are available, and degenerate hops are associated with symmetry.\footnote{We ignore accidental degeneracy throughout.}
We illustrate this point in Fig.~\ref{fig:intro_hop_symmetry}: each degenerate hop takes the composite to its image under a symmetry of the lattice (Fig.~\ref{fig:intro_hop_symmetry}(a)), so the same composite has fewer degenerate hops on a less symmetric lattice (Fig.~\ref{fig:intro_hop_symmetry}(b)), and a composite of irregular shape may have no degenerate hop at all (Fig.~\ref{fig:intro_hop_symmetry}(c)).
More precisely, only those lattice symmetries that map the composite pattern to a new one that differs from the initial one by only one particle can protect a degenerate hop. 
But what exactly is the requirement on the symmetries in order to achieve 2D rolling? We will study this question in Sec.~\ref{sec:theory}, where we will arrive at a criterion for 2D rollability expressed in terms of symmetry operators.
Guided by this criterion, in Sec.~\ref{sec:construction} we move on to construct all minimal settings that can host 2D-rollable 3e1h composites.
Here minimal means that every ingredient, including the lattice symmetries, orbitals and hoppings, is necessary for 2D rollability. Eventually, we find in total 14 minimal lattice models (readers interested only in the results may skip to Fig.~\ref{fig:hopping_models}), which can be considered as prototypes of systems that have the potential to reach a $T_c$ bounded only by the electron bandwidth. Up to this point, all the analysis is based on symmetry and is independent of the details of the Hamiltonian. 
We then ask whether a natural Hamiltonian can make the rollable composite the ground state, which requires a case-by-case analysis. 
	Therefore, we pick two of the 14 models. One is the triangle-kagome model (see Fig.~\ref{fig:hopping_models}), which is exactly the example shown in Fig.~\ref{fig:intro_rolling}(c), in which the electron orbitals form a triangular lattice and the hole orbitals the kagome lattice of its bond midpoints. The other is the AA-stacked triangle model (see Fig.~\ref{fig:hopping_models}), which is exactly the example described in Fig.~\ref{fig:intro_rolling}(d), in which the hole orbitals form the same triangular lattice as the electron orbitals.
For simplicity, we take the electrons and holes to be spinless, as in the spin- and valley-polarized quarter metal of rhombohedral graphene, where superconductivity has been observed~\cite{han_2025_chiral_sc}. 
In Sec.~\ref{sec:interaction_stabilization} we show that in both models the rollable 3e1h composite can be stabilized by the hole's kinetic energy. 
Once stabilized, the composites condense into a superconductor with $T_c$ being a fraction of the electrons' kinetic energy scale even when the interaction is extremely strong, as estimated in Sec.~\ref{sec:boson_tc}. 
Finally, in Sec.~\ref{sec:realization_gates}, we propose to realize this scenario in the dual-gate bilayer sketched in Fig.~\ref{fig:single_gate_geometry},
where the electron and hole orbitals are placed in two layers, so that the dual gate enables separately tunable electron and hole energies and, at the same time, screens the Coulomb interaction into the required short-range form. 
We will also test this proposal and the predicted scenario with exact numerics, which give a $T_c$ of a few percent of the electron hopping.

\section{Criteria of rolling}
\label{sec:theory}

In this section we consider what it takes for a generic composite with multiple particles to be 2D-rollable on a lattice. We will see that this kind of rollability is equivalent to a simple constraint on the symmetry of the lattice and the shape of the composite pattern.

\begin{figure}[!tb]
\centering
\includegraphics[width=\columnwidth]{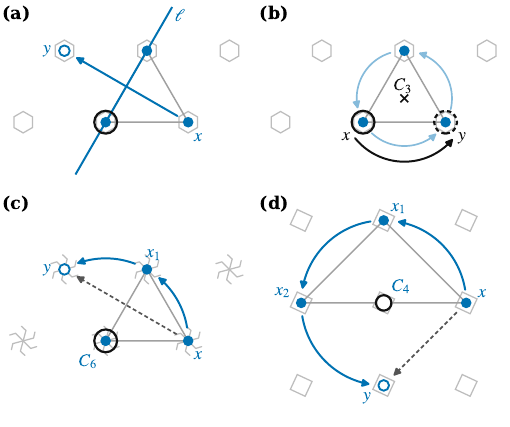}
\caption{
Examples of rolling operators. 
Naively, as hopping moves only one particle while keeping the rest untouched, it might appear that the rolling operator also needs to move one particle while keeping the pattern of the remaining particles unchanged, and thus has to be a symmetry of the pattern of the remainder, as are (a)~a reflection that carries an electron and (b)~a threefold rotation that carries the hole and permutes the electrons. However, in general, it does not need to  be such a symmetry. It can act on the whole pattern by moving a ``string'' of electrons. Examples with a string of length 2 and length 3 are given in (c) and (d), where blue solid arrows show how the rolling operator acts on the pattern. In these cases, while the rolling operator does not map the mover to its final position directly, it does map the whole pattern to a final state that is degenerate and differs from the initial one by only one electron, and thus still protects the degeneracy of the hop (dashed gray arrow).
}
\label{fig:strings}
\end{figure}

Rolling, by our definition above, is the movement through single-particle degenerate hops, i.e., the hops in which the initial and final composite patterns have the same energy. Without resorting to fine tuning, two patterns have the same energy only if a symmetry of the lattice maps one onto the other. Here the lattice includes its orbital structure, and its symmetries act on the orbitals as well as on positions. Let $C$ be the pattern of the composite on the lattice $L$, and let $C_{x\to y}$ be the pattern after one particle, which we call the \emph{mover}, hops from state $x$ to state $y$. This hop is therefore degenerate only if
\begin{equation}
PC=C_{x\to y}
\label{eq:rolling_condition}
\end{equation}
for some lattice symmetry $P$. We call 
$P$ a \emph{rolling operator} of $C$ [examples in Fig.~\ref{fig:strings}], as it generates one step of rolling.
We denote the set of all rolling operators of $C$ by $\mathcal G(C,L)$.

Using the language of rolling operators, the original question on rollability is essentially asking: For a composite $C$ to be 2D-rollable on the lattice $L$, what does its rolling operator set $\mathcal G(C,L)$ need to satisfy? 

As a preparation, we must identify which final patterns can be reached by a composite starting from an initial pattern $C_0$ through successive steps of rolling. After a first step to $C_1=P_1C_0$, the next step is generated by a rolling operator of $C_1$, which can be generally expressed as $P_1P_2P_1^{-1}$ with $P_2\in\mathcal G(C_0,L)$, because $P_2$ satisfies Eq.~\eqref{eq:rolling_condition} for $C_0$, with $x$ and $y$, exactly when $P_1P_2P_1^{-1}$ satisfies Eq.~\eqref{eq:rolling_condition} for $C_1$, with $P_1x$ and $P_1y$. The second step therefore  yields $C_2 = (P_1 P_2 P_1^{-1}) P_1 C_0 = P_1 P_2 C_0$. Generalizing this further, we find that a generic sequence of $k$ steps of rolling can be expressed as
\begin{equation}
C_k\;=\;P_1P_2\cdots P_kC_0,\qquad P_j\in\mathcal G(C_0,L).
\label{eq:rolling_sequence}
\end{equation}
Therefore, for a given initial pattern $C_0$, all reachable patterns by rolling are those that can be generated by a product of rolling operators of the \emph{initial} pattern $C_0$. The operators that carry $C_0$ to its reachable patterns, i.e., all products $P_1P_2\cdots P_k$ of elements of $\mathcal G(C_0,L)$, form a group, since the inverse of a rolling operator is also a rolling operator.
With this preparation, we now move on to address the key question with the rollability criterion.

\textbf{Rollability criterion.} \emph{A composite pattern $C_0$ is 2D-rollable on a lattice $L$ if and only if no point is left in place by all elements of $\mathcal G(C_0,L)$ and no line is mapped onto itself by all of them.}

\emph{Proof.} We first show that if $C_0$ is 2D-rollable, no point is left in place by all elements of $\mathcal G(C_0,L)$ and no line is mapped onto itself by all of them. If such a point or line existed, every product of these elements would also leave it in place or map it onto itself, and would thus preserve distances from it. By Eq.~\eqref{eq:rolling_sequence}, the reachable patterns would then stay within a bounded region around the point or within a strip along the line, so $C_0$ would not be 2D-rollable.

We now show the converse, that $C_0$ is 2D-rollable if no point is left in place by all elements of $\mathcal G(C_0,L)$ and no line is mapped onto itself by all of them. We prove its contrapositive. Suppose that $C_0$ is not 2D-rollable, so that its rolling is either confined to (i) a bounded region or (ii) a one-dimensional strip. Below we show that some point must then be left in place by all elements of $\mathcal G(C_0,L)$ in case (i), whereas some line must be mapped onto itself by all of them in case (ii). 

\begin{figure}[!tb]
\centering
\includegraphics[width=\columnwidth]{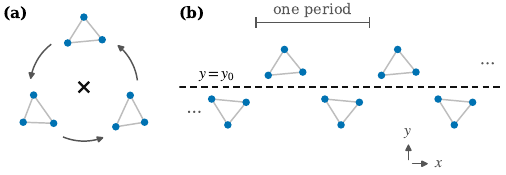}
\caption{The two ways in which rolling can fail to be two-dimensional. (a)~Rolling confined to a bounded region: every rolling operator permutes the finitely many reachable patterns (gray arrows) and thus leaves their center of mass (cross) in place. (b)~Rolling confined to one dimension: the reachable patterns repeat periodically along $x$, so the height $y_0$ of their center of mass is well-defined, and every rolling operator maps the line $y=y_0$ (dashed) onto itself.}

\label{fig:proof_cases}
\end{figure}

Let $\mathcal S$ be the set of reachable patterns, i.e., the images of $C_0$ under the group of products in Eq.~\eqref{eq:rolling_sequence}. Multiplying all elements of a group by one of them only reorders them, so every element $Q$ of $\mathcal G(C_0,L)$ maps $\mathcal S$ onto itself. As a result, for case (i) where $\mathcal S$ is bounded, the center of mass of $\mathcal S$ is invariant under $Q$. This point is therefore left in place by all elements of $\mathcal G(C_0,L)$.
For case (ii) where rolling is confined to 1D, let $x$ be the direction of unbounded rolling and $y$ the perpendicular direction where rolling is confined. $\mathcal S$ extends to infinity along $x$ and is periodic along it due to the presence of the lattice,
so the $y$ coordinate $y_0$ of its center of mass is well-defined [Fig.~\ref{fig:proof_cases}(b)]. Since $Q$ maps $\mathcal S$ onto itself, it preserves the ``height'' of the center of mass $y_0$. Therefore, the line $y=y_0$ is mapped onto itself by all elements of $\mathcal G(C_0,L)$.
$\square$

\section{Constructing minimal lattice models for 2D-rollable 3e1h composites}%
\label{sec:construction}

The rollability criterion reduces 2D rolling to a condition on the symmetry of the lattice and the shape of the composite pattern. Since this condition involves only symmetry and shape, it can be used not only to test whether a given lattice model allows rollable composites but also to derive the lattices that host a 2D-rollable composite. In this section, we focus on 3e1h composites, and we would like to find all the lattice models that host a 2D-rollable 3e1h composite, i.e., all lattices with their orbital arrangement and the hopping terms of the Hamiltonian, along with the rollable 3e1h pattern each lattice model hosts. 

However, there are infinitely many such models, even for a given shape of the composite. This is because, once a composite is 2D-rollable on a lattice model, adding further symmetries, orbitals or hoppings that respect the symmetries already present only adds degenerate hops without hurting existing ones, so the composite remains 2D-rollable. Listing all of them would therefore be neither possible nor informative. What is fundamentally interesting is rather what 2D rolling needs at the least, i.e., the \emph{minimal} lattice models, in which no ingredient (symmetry, orbital, hopping) can be removed without losing 2D rolling. In this way, all other lattice models with a 2D-rollable composite can be viewed as a derivative of some of them with higher symmetry, more orbitals or more hopping terms. 

Therefore, in this section, we aim to construct 2D-rollable 3e1h patterns and their minimal lattice models, minimal in every sense, with minimal symmetry, minimal orbitals and minimal hoppings. This construction will be based on the general criterion we found in the last section. Our plan is to construct these models in three steps, each of which minimizes one of the three ingredients. First, we find the minimal symmetry, in the form of a minimal set of rolling operators. Starting from a 3e1h composite with arbitrary positions and states, we add rolling operators one at a time, each of which restricts the shape of the pattern and must be a symmetry of the lattice, and stop as soon as the criterion is satisfied. The results are configurations, i.e., combinations of a pattern and a set of its rolling operators, that give 2D rolling and from which no operator can be removed without losing it. We call each of them a ``minimal 2D-rollable configuration''. Second, we find the minimal orbitals. For each configuration, we let the composite roll, i.e., apply the rolling operators to the initial composite again and again, and the orbitals that it visits along the way form the minimal lattice of the configuration, since every lattice hosting the configuration must contain them. Because the rolling operators must be symmetries of this lattice, the lattice in turn allows only a few shapes of the pattern, so we construct it right after each configuration. Up to this point, we assume that the Hamiltonian contains the hopping term of every degenerate hop, so that every rolling operator generates a step of rolling%
. Third, %
 we find the minimal hoppings: we will determine which of these hoppings 2D rolling actually needs, which turns each minimal lattice into one or a few minimal lattice models. The resulting models are collected in Fig.~\ref{fig:hopping_models} for readers interested only in the results.

Before beginning our analysis we note that we will always assume the three electrons lie at distinct positions, whereas the hole is allowed to be placed anywhere, even on top of an electron. While two electrons may occupy different orbitals or spin states at the same position, in the systems of interest here, with strong e-e repulsion, it is usually energetically costly to have two electrons sitting at zero in-plane distance. Therefore, we defer the analysis of these special situations to elsewhere.

To keep the construction as general as possible, we allow the electrons to occupy states that differ in spin or local orbital character. However, a symmetry operator that maps one electron to another must also be a symmetry of the lattice. It therefore requires the spin and orbital states of these two electrons to be related by that symmetry. Electrons not related by the required symmetry operators need not occupy symmetry-related states. To keep track of these constraints as the construction develops, we use the following \hypertarget{par:electron_color_convention}{\textbf{color convention}} for the electron symbols: electrons related by the operators are drawn in the same color, while those not required to be related are colored differently. In the following derivation, when we add to the system a symmetry operator that relates previously unrelated electrons, we will identify their colors. The final colors then make clear which electrons must occupy symmetry-related spin and orbital states,\footnote{For the spinless fermions considered in this paper, this color convention concerns only the orbital degrees of freedom.} whereas the others need not. We use this color convention throughout the paper. 

We now construct the minimal configurations. The constraint that a rolling operator places on the pattern depends on which particle it moves. We therefore call a rolling operator a \emph{hole rolling operator} if its mover is the hole, and an \emph{electron rolling operator} if its mover is an electron. By Eq.~\eqref{eq:rolling_condition}, a hole rolling operator maps the three-electron pattern onto itself and the hole to a new state, so it is a symmetry of the three-electron pattern. An electron rolling operator must map the hole onto itself, so it leaves the hole position $h$ in place. Both kinds are therefore needed, since with only one of them either the three electrons or the hole would stay in place. At least one hole rolling operator must also move the hole to a new position, rather than only change its state at the same position, since otherwise every rolling operator would leave $h$ in place. Therefore, our construction of lattice models should start by introducing the first electron and hole rolling operators.

\begin{figure}[!tb]
\centering
\includegraphics[width=\columnwidth]{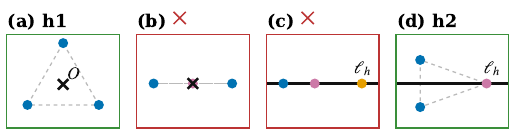}
\caption[Hole rolling operators]{A hole rolling operator maps the three-electron pattern onto itself, which is possible only when the operator is one of the following four cases: 
	 (a) a three-fold rotation, (b) a two-fold rotation, and (c),(d) a reflection. Black crosses mark the rotation centers, the one in (a) being the center $O$ of the triangle. Cases (b) and (c) are excluded (marked by red crosses) because in these cases, all three electrons are on one line, and therefore none of the electrons can hop off this line without deformation. Panels (a) and (d) illustrate the cases h1 and h2 in the main text. %
	 }

\label{fig:hole_operator_cases}
\end{figure}

Let us now start the construction by adding the first hole rolling operator and see how that would constrain the configuration. 
First, if there is a rotation, there is only one way to arrange it, which is shown in Fig.~\ref{fig:hole_operator_cases}(a), where the rotation is $C_3$ and the electrons form an equilateral triangle, so that the rotation permutes electrons without changing the whole 3e pattern. A $C_2$ rotation that exchanges two electrons while preserving the third is excluded [see Fig.~\ref{fig:hole_operator_cases}(b)], since it requires the three electrons to lie on one line. This is not allowed, because electrons on one line can never hop off it, as such a hop would inevitably change the shape of the pattern.
Next, how would a mirror constrain the configuration? A reflection leaves in place the electrons on its mirror line and exchanges the others in pairs. The case shown in Fig.~\ref{fig:hole_operator_cases}(c) where three electrons all sit on the mirror is excluded as this would restrict the rolling to 1D. Therefore, when there is a mirror, we are also left with only one way to arrange it:
pass through one electron while perpendicularly bisecting the other two [Fig.~\ref{fig:hole_operator_cases}(d)]. 
Thus, the hole rolling operator set $\mathcal G_h$ can be classified into two cases:
\begin{itemize}
\item[\textbf{(h1)}] $\mathcal G_h$ contains at least a $C_3$ rotation. The triangle is equilateral, and the hole lies away from its center $O$ [Fig.~\ref{fig:hole_operator_cases}(a)]. The hole cannot sit at $O$, since otherwise none of the hole rolling operators would be able to move the hole to a new position, as they must keep the 3e triangle invariant.
\item[\textbf{(h2)}] $\mathcal G_h$ does not contain any rotation. A hole rolling operator is then the reflection $P_h$ in a symmetry axis $\ell_h$ of the triangle, with the hole off $\ell_h$ [Fig.~\ref{fig:hole_operator_cases}(d)]. Note that in this case, it is the only operator allowed because adding a second mirror $P_h'$ would combine with $P_h$ into a rotation in $\mathcal G_h$, which is not allowed in this case and belongs to case h1. As the only hole rolling operator, $P_h$ must move the hole to a new position, so the hole must lie off $\ell_h$.
\end{itemize}

Next, we add an electron rolling operator and see how it constrains the configuration. Since it leaves $h$ in place, it is either a rotation about $h$ or a reflection in a line through $h$, and it maps the three electrons onto the three electrons after the hop, two of which have not moved. 
	We note that, while the mover moves directly from its initial to its final site in a hop, an electron rolling operator does not have to carry the mover electron straight to its final site. It can carry the mover onto a second electron, and the second onto a third, and so forth, as long as the last one lands on the final site whereas the sites of all intermediate electrons on the ``string'' remain occupied [Fig.~\ref{fig:strings}(c),(d)]. 
	As there are three electrons, the string of a rolling operator can cover one, two or three electrons. We call it the length of the string. 

\begin{figure}[!tb]
\centering
\includegraphics[width=\columnwidth]{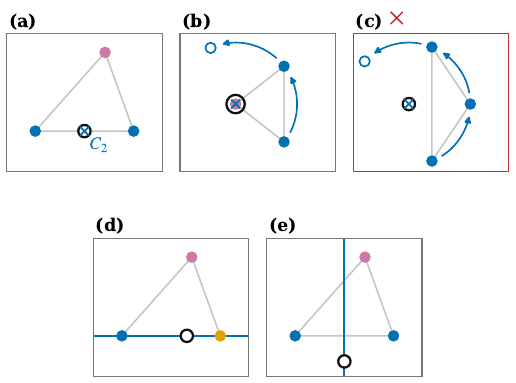}
\caption{The possible electron rolling operators. An electron rolling operator has to leave the hole (black circle) in place, so it is a rotation about the hole or a reflection in a line through it. (a)--(c)~A rotation can carry a string of one, two or three electrons. (a)~For a string of one electron, the rotation must be the $C_2$ rotation about the midpoint of the other two electrons, where the hole sits. (b)~For a string of two electrons, both the hole and the third electron have to sit on the rotation center, equidistant from the other two. Blue arrows describe how this rotation moves the pattern. (c)~For a string of three electrons, the hole is the one point equidistant from all three electrons. As a result, this hole has no other degenerate position to hop to. Therefore, this case is excluded. (d),(e)~A reflection should preserve a two-electron pattern while moving the third electron, so it can either (d)~pass through these two electrons or (e)~perpendicularly bisect them. In both cases, the hole has to sit on this mirror line. 
}

\label{fig:electron_operators}
\end{figure}

First, if the operator is a rotation, its string can have any of these lengths. (i) For a rotation with a string of length 1e, where the rotation moves the mover directly to its final site, it has to be a $C_2$ rotation about the midpoint of the other two electrons, as shown in Fig.~\ref{fig:electron_operators}(a), and the hole has to also sit on the midpoint.
(ii) For a rotation with a string of length 2e, it leaves the third electron in place, so this electron sits at the center, on top of the hole, at the same distance from the other two
 [Fig.~\ref{fig:electron_operators}(b)]. 
(iii) For a rotation with a string of length 3e, all electrons must lie at the same distance from the hole, as shown in Fig.~\ref{fig:electron_operators}(c). However, this case never works, because only one point in the plane is at the same distance from three electrons that do not lie on one line, so the hole has no other position to hop to.
Next, if the operator is a reflection, its string can only be length 1e.
Such a reflection has to either pass through the two electrons other than the mover [Fig.~\ref{fig:electron_operators}(d)], or 
perpendicularly bisect them [Fig.~\ref{fig:electron_operators}(e)]. In both cases the hole must sit on the mirror line.

\begin{figure}[!tb]
\centering
\includegraphics[width=\columnwidth]{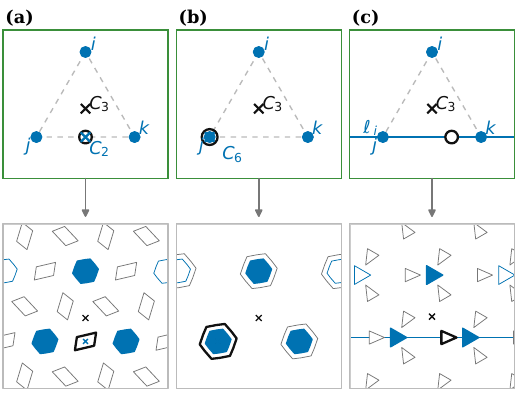}
\caption{The three minimal configurations of case h1 (top) and the minimal lattices hosting (and derived from) each of them (bottom). The $C_3$ rotation about the center $O$ of the electron triangle (black) leaves only $O$ in place and maps no line onto itself, so 2D rolling can be achieved by simply adding a single electron rolling operator (blue) that moves $O$. There are three electron rolling operators that meet this goal: Figs.~\ref{fig:electron_operators}(a), (b) and (d), as their centers and mirror line lie away from $O$. These three cases are shown in panels (a), (b) and (c), respectively. (a)~Adding the $C_2$ rotation about the midpoint of $j,k$, (b)~Adding the $C_6$ rotation about $j$, on which the hole sits, and (c)~Adding the reflection in $\ell_i$. 
Rolling each configuration gives its minimal lattice, formed by the orbitals the composite visits. Here and below, %
the notation $\ell_x$ ($x=i,j,k$) denotes the line through the two electrons other than $x$ and $\ell_x'$ denotes their perpendicular bisector, the mirror lines of the two reflections whose mover is $x$. In the lattices, filled symbols are the orbitals occupied by the composite, with the hole outlined in black, open symbols are the other orbitals (gray for the hole), the shape of each symbol has the necessary symmetry of its site, and crosses mark the rotation centers.
In (c), the hole h can sit anywhere on $\ell_i$.}

\label{fig:h1_configurations}
\end{figure}

Having listed the possible electron rolling operators [Fig.~\ref{fig:electron_operators}], we now return to the two cases h1 and h2 and add these operators to each of them, starting with h1. Let us first see what the criterion requires. The $C_3$ rotation leaves only the center $O$ of the triangle in place and maps no line onto itself, so the criterion only requires one electron rolling operator that moves the center $O$. Adding the operators of Figs.~\ref{fig:electron_operators}(a), (b) and (d), we find that each of them works without needing more operators (see Figs.~\ref{fig:h1_configurations}(a)--(c)), since their rotation centers and mirror line lie away from the center $O$. 
However, the reflection of Fig.~\ref{fig:electron_operators}(e) does not work, since in the equilateral triangle its mirror line passes through the center $O$. We note that this case cannot be rescued by adding more operators, as adding more hole rolling operators or more reflections of this kind leaves the center $O$ in place as well, whereas adding an operator of Figs.~\ref{fig:electron_operators}(a), (b) or (d) works but is not a ``minimal'' configuration, as the previous reflection of Fig.~\ref{fig:electron_operators}(e) is then unnecessary for rollability. Therefore, for case h1, we conclude that there are three minimal configurations, as shown in Fig.~\ref{fig:h1_configurations}. To get the minimal lattice that can host the rolling, where ``minimal'' is in the sense that all orbitals (sites) are needed for rolling whereas symmetry is as low as possible, we simply roll each configuration by acting with the rolling operators on it. The orbitals that the composite visits in the rolling form its minimal lattice, shown below each configuration in Fig.~\ref{fig:h1_configurations}.

\begin{figure}[!tb]
\centering
\includegraphics[width=\columnwidth]{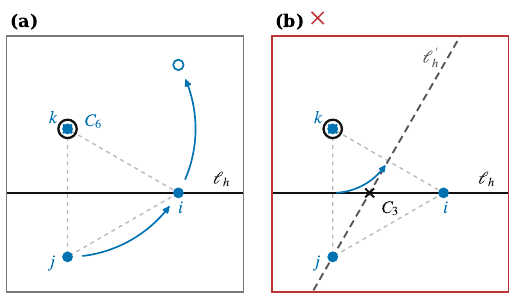}
\caption{Why the rotation about an electron on the hole [the operator in Fig.~\ref{fig:electron_operators}(b)] is excluded from case h2. (a)~Since the hole lies off $\ell_h$, it sits on one of the two electrons that $P_h$ exchanges, here $k$, which is equidistant from the other two only if the triangle is equilateral. This restricts the rotation to a $C_6$ rotation. (b)~The $C_6$ rotation carries $\ell_h$ onto the symmetry axis $\ell_h'$ through $j$ (blue arc), so the reflection in $\ell_h'$ is a lattice symmetry as well. The reflections in $\ell_h$ and $\ell_h'$ meet at the electron triangle's center $O$ at $60^\circ$ and generate the $C_3$ rotation about $O$ in the lattice symmetry group. This $C_3$ moves only the hole, without changing the electron pattern. Therefore, it is a hole rolling operator. As a result, this case should be excluded from h2, and should repeat the cases already considered in h1 or their higher-symmetry variations.}

\label{fig:h2_rotation_excluded}
\end{figure}

In case h2, we add the electron rolling operators in the same way. $P_h$ leaves every point of $\ell_h$ in place and maps every line perpendicular to $\ell_h$ onto itself. By the criterion, each of these points and each of these lines must therefore be moved by at least one electron rolling operator. We go through the valid types in Fig.~\ref{fig:electron_operators}, and whenever an operator does not work by itself, we consider adding a second one. %
However, the rotation about an electron on the hole [Fig.~\ref{fig:electron_operators}(b)] is excluded from case h2 based on the following reasoning:
Such a rotation needs the hole on an electron at the same distance from the other two. As the hole lies off $\ell_h$,  $\ell_h$ has to pass through one of the two electrons away from the hole, which then forces the 3e triangle to be equilateral and the rotation to be a $C_6$ rotation [Fig.~\ref{fig:h2_rotation_excluded}(a)]. With both $P_h$ and this $C_6$ in the lattice symmetry group, the $C_3$ about the center $O$ of the triangle also becomes a lattice symmetry as shown in Fig.~\ref{fig:h2_rotation_excluded}(b). This $C_3$ rotation keeps the three-electron pattern invariant
	and moves the hole, so by Eq.~\eqref{eq:rolling_condition} it is a hole rolling operator, which case h2 does not allow. Therefore, below we only need to pick from three types of electron rolling operators: Fig.~\ref{fig:electron_operators}(a), (d) and (e).

\begin{figure}[!tb]
\centering
\includegraphics[width=\columnwidth]{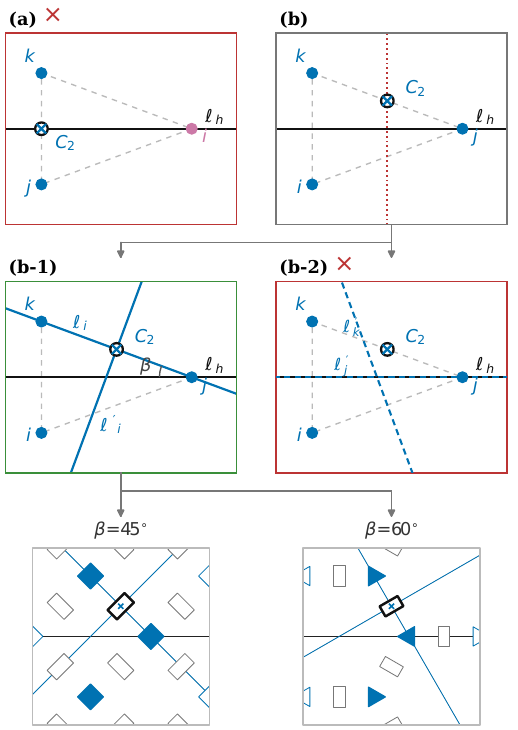}
\caption{Case h2 with the $C_2$ rotation about the hole [Fig.~\ref{fig:electron_operators}(a)] as an electron rolling operator with the electron labeled by $i$ being its mover. 
	(a)~If the hole sits at the midpoint of the two electrons that $P_h$ exchanges, it lies on $\ell_h$, which is excluded. (b)~If it sits at the midpoint of the electron on $\ell_h$ and another electron, the $C_2$ rotation and $P_h$ both map the red dotted line onto itself, which confines the rolling to this line. To achieve 2D rolling we need to add another mirror. (b-1)~Adding reflections in $\ell_i$ and $\ell_i'$. The two mirrors come together due to the presence of the $C_2$ rotation. This is a valid 2D-rollable configuration. 	Here,
	$\beta$ is the angle between $\ell_h$ and $\ell_i$. This configuration (b-1) can be hosted by two minimal lattices that are generated by the configuration at $\beta=45^\circ$ and $60^\circ$, drawn as in Fig.~\ref{fig:h1_configurations}. (b-2)~It is impossible to add any other reflections to the electron rolling operators, because the added mirror lines cannot pass through the hole. 
}

\label{fig:h2_c2}
\end{figure}

We start by adding to h2 the $C_2$ rotation [Fig.~\ref{fig:electron_operators}(a)], which is centered at the midpoint of two electrons with the hole sitting on it, so there are two inequivalent placements, shown in Fig.~\ref{fig:h2_c2}(a) and (b). The placement in (a) does not work, as the hole would have to sit on $\ell_h$, which is banned in h2. Therefore the placement must be like in Fig.~\ref{fig:h2_c2}(b). However, this configuration does not work by itself, since $C_2$ maps every line through the hole onto itself, including the one perpendicular to $\ell_h$ [red dotted line in Fig.~\ref{fig:h2_c2}(b)], so the rolling is confined to this line. As a remedy, we therefore consider adding a second electron rolling operator: Another $C_2$ rotation about the hole would be the same one, so instead of that, we have to add a reflection in a line through the hole. Among the two possible types of mirror lines in Figs.~\ref{fig:electron_operators}(d) and (e), only two pass through the hole, i.e., the line $\ell_i$ through the two electrons and their perpendicular bisector $\ell_i'$ (see Fig.~\ref{fig:h2_c2}(b-1)), whereas all others miss the hole (see Fig.~\ref{fig:h2_c2}(b-2)) and are therefore excluded.
Interestingly, these two mirror lines always come together: in the presence of one mirror line, say, $\ell_i$, the other mirror line $\ell_i'$ will automatically become a lattice symmetry due to the $C_2$ rotation. Then, as $\ell_i'$ also satisfies Eq.~\eqref{eq:rolling_condition}, it must be a rolling operator. Therefore, we conclude that in this case there is only one valid configuration shown in Fig.~\ref{fig:h2_c2}(b-1), where the electron rolling operator set contains two mirrors and a $C_2$. Now no point is left in place by all these rolling operators and no line is mapped onto itself by all of them, so this configuration gives 2D rolling. This configuration has one free parameter, the angle $\beta$ between $\ell_h$ and $\ell_i$ [Fig.~\ref{fig:h2_c2}(b-1)]. 
	Since these two reflections generate an angle-$2\beta$ rotation symmetry in the lattice symmetry group, which is restricted to order 2, 3, 4 or 6, the angle $\beta$ is restricted to the following discrete values,
	 $\beta=30^\circ$, $45^\circ$ or $60^\circ$ ($90^\circ$ is impossible as the electrons need to form a triangle). However, we find that $\beta=30^\circ$ is excluded from case h2, because the triangle is then equilateral and $\ell_i'$ becomes the symmetry axis through $i$, so the reflection in $\ell_i'$ and $P_h$ generate the $C_3$ rotation about $O$, which is a hole rolling operator as in Fig.~\ref{fig:h2_rotation_excluded}(b). Therefore, only $\beta=45^\circ$ and $60^\circ$ survive, and they give the two minimal lattices at the bottom of Fig.~\ref{fig:h2_c2}.

\begin{figure*}[t]
\centering
\includegraphics[width=\textwidth]{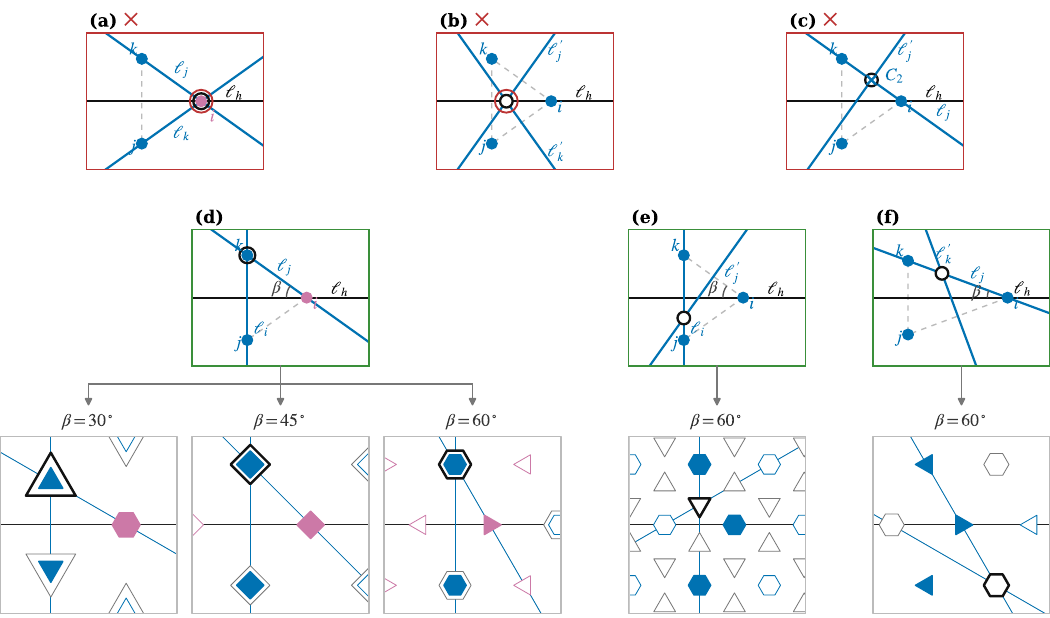}
\caption{Case h2 where the electron rolling operators contain only reflections [see Figs.~\ref{fig:electron_operators}(d),(e)]. A single reflection is not enough, because its mirror line crosses $\ell_h$, and both this reflection and $P_h$ leave the crossing point in place, so the composite stays within a bounded region. Therefore, we need at least two reflections. Since each of them leaves the hole in place, their mirror lines cross at the hole. We label the electron on $\ell_h$ as $i$ and the two electrons that $P_h$ exchanges as $j$ and $k$. The mirror line of each reflection can then be $\ell_i$, $\ell_j$, $\ell_k$, $\ell_j'$ or $\ell_k'$, whereas $\ell_i'$ is $\ell_h$ itself. Up to the mirror image under $P_h$, there are six inequivalent ways to choose two of these lines, shown in panels (a)--(f). The three choices in the top row are excluded (red crosses). In (a), $\ell_j$ and $\ell_k$ cross at the electron $i$, and in (b), $\ell_j'$ and $\ell_k'$ also cross on $\ell_h$, so in both cases the hole is forced to sit on $\ell_h$, which is not allowed in case h2. In (c), $\ell_j$ and $\ell_j'$ are perpendicular to each other, so the two reflections generate the $C_2$ rotation about the midpoint of $i$ and $k$, and this case repeats the configuration of Fig.~\ref{fig:h2_c2}(b-1). The three choices in the middle row are valid 2D-rollable configurations: (d)~$\ell_i$ and $\ell_j$, which cross at the electron $k$, so the hole sits on top of $k$, (e)~$\ell_i$ and $\ell_j'$, and (f)~$\ell_j$ and $\ell_k'$. In each of them, the two mirror lines cross at the hole at an oblique angle, so no point is left in place and no line is mapped onto itself by all the rolling operators, which gives 2D rolling.
	They are drawn at a generic angle $\beta$ between $\ell_h$ and the two sides of the triangle through $i$. The bottom row shows the minimal lattices generated by these configurations at the values of $\beta$ that the lattice allows (see analysis in the main text), $\beta=30^\circ$, $45^\circ$ and $60^\circ$ for (d), and $\beta=60^\circ$ for (e) and (f). 
}

\label{fig:h2_reflections}
\end{figure*}

\begin{figure*}[t]
\centering
\includegraphics[width=\textwidth]{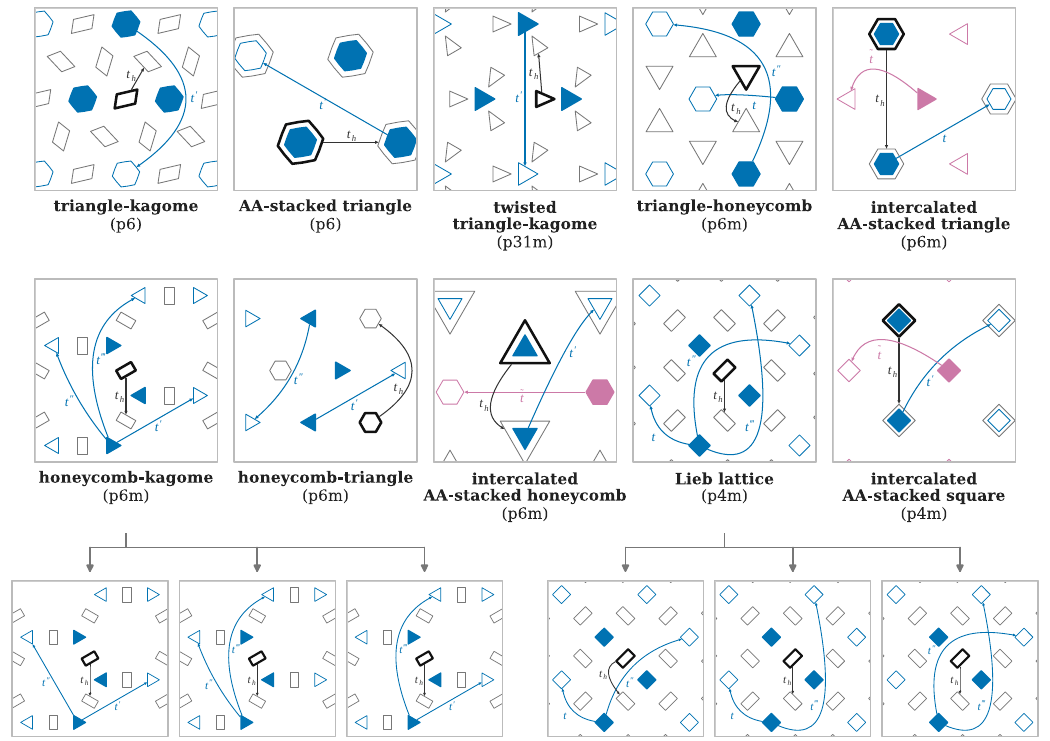}
\caption{The fourteen minimal lattice models. Top two rows: the ten minimal lattices, with the hoppings corresponding to the rolling operators of their configurations as arrows in the color of the mover (an electron hopping between first, second, third or fourth neighbors on the lattice of its orbitals is $t$, $t'$, $t''$ or $t'''$, with a tilde for the purple orbitals, and the hole hopping is $t_h$). Each arrow also represents its Hermitian conjugate and its symmetry-related copies. 
		The models are named based on the lattice of the (blue) electron orbitals, followed by that of the hole orbitals. %
		In the intercalated lattices, a second electron orbital (purple), inequivalent to the blue one, sits at the plaquette centers of the lattice of the blue electron orbitals. 
		For each case, we add the symbol of its wallpaper group in parentheses. Triangle-kagome, AA-stacked triangle and twisted triangle-kagome come from Fig.~\ref{fig:h1_configurations}(a)--(c), the Lieb lattice and honeycomb-kagome from Fig.~\ref{fig:h2_c2}(b-1) at $\beta=45^\circ$ and $60^\circ$, intercalated AA-stacked honeycomb, intercalated AA-stacked square and intercalated AA-stacked triangle from Fig.~\ref{fig:h2_reflections}(d) at $\beta=30^\circ$, $45^\circ$ and $60^\circ$, and triangle-honeycomb and honeycomb-triangle from Figs.~\ref{fig:h2_reflections}(e) and (f). Eight of these are already minimal lattice models. Bottom row: honeycomb-kagome and the Lieb lattice each give three, which keep any two of the three electron hoppings generated by the $C_2$ rotation and the reflections in $\ell_i$ and $\ell_i'$, namely $t'''$, $t'$ and $t''$ in honeycomb-kagome and $t'''$, $t''$ and $t$ in the Lieb lattice. 
}
\label{fig:hopping_models}
\end{figure*}

Next, we consider the case where the electron rolling operators contain no rotation, only reflections [Figs.~\ref{fig:electron_operators}(d),(e)]. A single reflection does not work by itself, because its mirror line crosses $\ell_h$ and both it and $P_h$ leave the crossing point in place, so the composite stays within a bounded region. We therefore need at least two reflections in the electron rolling operator set, whose mirror lines cross at the hole. We label the electron on $\ell_h$ as $i$ and the two electrons that $P_h$ exchanges as $j,k$. A mirror line through the hole can then be the line $\ell_i$ through $j,k$, the line $\ell_j$ through $i,k$ or $\ell_k$ through $i,j$, or the perpendicular bisector $\ell_j'$ of $i,k$ or $\ell_k'$ of $i,j$, whereas the perpendicular bisector of $j,k$ is $\ell_h$ itself.
	Up to the mirror image under $P_h$, 
	there are six inequivalent ways to choose two lines, as shown in Fig.~\ref{fig:h2_reflections}(a)--(f).

Three of the six arrangements, Figs.~\ref{fig:h2_reflections}(a),(b) and (c), are excluded because in (a) and (b) the hole is forced to sit on $\ell_h$, whereas (c) repeats Fig.~\ref{fig:h2_c2}(b-1).

In the remaining three, Figs.~\ref{fig:h2_reflections}(d)--(f), the two mirror lines cross at the hole at an oblique angle. The two reflections then leave only the hole in place, which $P_h$ moves, and map no line onto itself together, so these configurations give 2D rolling, and a third reflection is unnecessary. Each of these three valid cases again has the free parameter $\beta$, the angle between $\ell_h$ and the sides through $i$, which the lattice restricts to $30^\circ$, $45^\circ$ or $60^\circ$ as for Fig.~\ref{fig:h2_c2}(b-1). In Fig.~\ref{fig:h2_reflections}(d), all three values are allowed and give the 
three lattices in Fig.~\ref{fig:h2_reflections}. 
In Figs.~\ref{fig:h2_reflections}(e) and (f), we find that only $\beta=60^\circ$ survives, which gives the last two lattices in Fig.~\ref{fig:h2_reflections}. The other two values are excluded for the following reasons. In Fig.~\ref{fig:h2_reflections}(e), at $\beta=45^\circ$ it is not hard to check that $\ell_j'$ passes through the midpoint of $j,k$, so the hole is forced to sit on $\ell_h$, which is banned;
at $\beta=30^\circ$, $\ell_j'$ becomes the symmetry axis through $j$, which generates a $C_3$ rotation in the hole rolling operator set as in the $\beta=30^\circ$ case of Fig.~\ref{fig:h2_c2}(b-1), so this value is excluded from h2. In Fig.~\ref{fig:h2_reflections}(f), at $\beta=45^\circ$, $\ell_k'$ becomes parallel to $\ell_j$, leaving no place for the hole, and at $\beta=30^\circ$, $\ell_k'$ becomes the symmetry axis through $k$, which again generates a $C_3$ hole rolling operator, so this value is excluded from case h2.

With these, we complete the symmetry-based construction of the lattice models. In total, we find seven minimal 2D-rollable configurations, three in case h1 [Fig.~\ref{fig:h1_configurations}] and four in case h2 [Figs.~\ref{fig:h2_c2}(b-1) and \ref{fig:h2_reflections}(d)--(f)], which yield ten minimal lattices. Recall that each configuration consists of a composite pattern and its minimal set of rolling operators supporting 2D rolling. Finally, to complete the construction of minimal lattice models, we need to work out the minimal set of hoppings for each configuration, where ``minimal'' means necessary for 2D rolling. As we discussed, each rolling operator translates directly into the hopping of the degenerate hop it generates, together with its symmetry-related copies. We collect the ten lattices with these hoppings in the first two rows of Fig.~\ref{fig:hopping_models}, where we also name them after the lattices formed by their electron and hole orbitals, as detailed in the caption. 

Are these already the ``minimal'' lattice models with ``minimal'' sets of hoppings? Almost, but not strictly. It might appear that, as the rolling operators in the configurations we worked out are already ``minimal'', all hoppings generated by these rolling operators are already the ``necessary'' ones. However, it turns out that this is not always the case. 
	Namely, in both the Lieb lattice and the honeycomb-kagome lattice (see the panel names in Fig.~\ref{fig:hopping_models}) we find that, while symmetry makes three inequivalent hops degenerate, any two of their three hoppings are already enough to achieve 2D rollability.

This is not accidental. Tracing the Lieb lattice and the honeycomb-kagome lattice back, we find that both come from the configuration of Fig.~\ref{fig:h2_c2}(b-1).
There, the reflection in $\ell_i'$ was not added by us but came automatically with the $C_2$ rotation and the reflection in $\ell_i$, and in fact any two of these three operators generate the third. The symmetry thus forces all three to be present, but any two of them already satisfy the criterion together with $P_h$, and the hop of the third can be replaced by two successive hops generated by the other two, so its hopping is unnecessary.
In comparison, in the other configurations, no rolling operator comes automatically, and removing any one of the rolling operators violates the criterion, as we saw while constructing them, so the hopping generated by each of them is necessary. 

As a result, we arrive at the fourteen minimal lattice models presented in Fig.~\ref{fig:hopping_models}, where each of the Lieb lattice and the honeycomb-kagome lattice gives three models, each keeping two of its three electron hoppings, whereas each of the other eight lattices yields only one model. In each of them, the 3e1h composite of the corresponding configuration rolls in two dimensions, but whether composites with these patterns can be stabilized energetically has not been determined so far. We will study this problem in the next section.

\section{Stabilizing a rollable composite}
\label{sec:interaction_stabilization}

In the last section we constructed noninteracting lattice models that host rollable 3e1h composites.
However, we need to have these rollable composites stabilized in the first place, which means we need these rollable composites to be energetically more favorable than other competitors, such as 2e1h, 4e1h, neutral 1e1h, or composites containing several holes.
In this section we study how to stabilize the rollable composite.

An intuitive picture is that interactions alone stabilize the rollable composite whereas all degenerate hops only take on the role of moving the composite. To see how it might work, we write down a general form of the interacting Hamiltonian in these models:
\begin{equation}
\begin{aligned}
H={}&-\sum_{\langle ij\rangle}t_{ij}\,c_i^\dagger c_j-t_h\sum_{\langle ab\rangle}h_a^\dagger h_b+\Delta\sum_a n^h_a\\
&+\sum_{i<j}U_{r_{ij}}n_in_j-\sum_{i,a}V_{r_{ia}}n_in^h_a+\sum_{a<b}W_{r_{ab}}n^h_an^h_b.
\end{aligned}
\label{eq:general_hamiltonian}
\end{equation}
Here $\Delta$ is the energy cost of a hole, i.e., the hole gap, with a free electron at zero energy, and $U_r$, $V_r$, $W_r$ denote the electron-electron repulsion, the electron-hole attraction and the hole-hole repulsion between two orbitals at the in-plane distance $r$, in units of the lattice constant $a$ of the electron orbitals, respectively. The operators $c_i$ and $h_a$ annihilate an electron in the electron orbital $i$ and a hole in the hole orbital $a$, $n_i=c_i^\dagger c_i$ and $n^h_a=h_a^\dagger h_a$, and the two hopping sums run over the hops of the lattice model [Fig.~\ref{fig:hopping_models}] and their Hermitian conjugates. 
All these three kinds of interactions play crucial roles in stabilizing the rollable composite. The electron-hole interaction $V$ takes on the role of holding the three electrons around the hole, whereas the electron-electron repulsion $U$ and the hole-hole repulsion $W$ play the role of keeping the composite to exactly 3e1h: Namely,
$U$ keeps the fourth electron from joining the composite while $W$ keeps the second hole from joining the composite. 

It still requires a case-by-case analysis to see whether such an interaction is sufficient to stabilize the rollable composite in a given model.
The answer depends on the model. With some simple algebra, one can show that in the triangle-kagome model a suitably designed short-range interaction can stabilize the rollable composite in some regime, while in the AA-stacked triangle model the interaction alone can never achieve this.
 However, we do not really want to rely on the interaction alone to select out the rollable composite, because (a) this way depends so strongly on the model details and parameter regime, such as the balance between the electron-electron repulsion and the electron-hole attraction, and (b) it immediately rules out too many possibilities, such as the whole AA-stacked triangle model, which makes this scenario relatively restrictive. So we ask, is there another way to stabilize the rollable 3e1h composite?

\begin{figure}[!tb]
\centering
\includegraphics[width=\columnwidth]{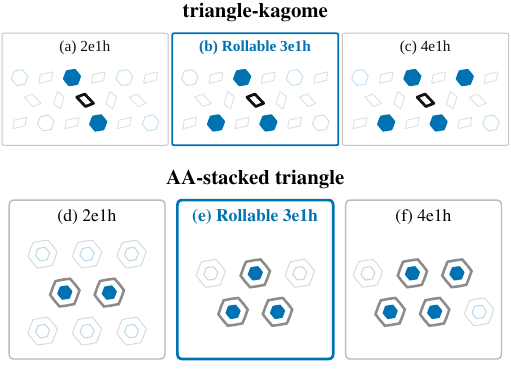}
\caption{The rollable composite and its closest competitors, the 2e1h and 4e1h composites, in the triangle-kagome model, (a)--(c), and the AA-stacked triangle model, (d)--(f). Filled blue polygons and the thick black polygon denote occupied electron and hole orbitals. Electron/hole orbitals with semitransparent open polygons are unoccupied. 
		In the AA-stacked triangle model, the hole is spread over the hole orbitals on top of the electrons, drawn as thick gray polygons, with $V_0\gg t_h$ and $t'=0$, so that electrons are on definite sites whereas the hole is in the bonding state.
		}
\label{fig:hardest_competitors}
\end{figure}

\begin{figure}[!tb]
\centering
\includegraphics[width=\columnwidth]{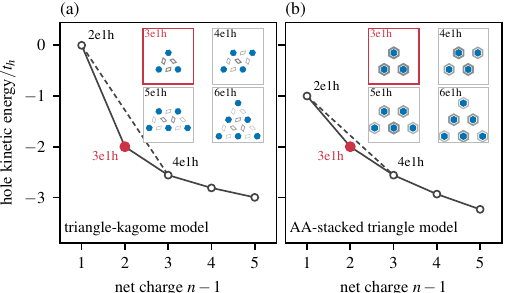}
\caption{Kinetic energy of the hole in the $n$e1h composite of (a) the triangle-kagome model and (b) the AA-stacked triangle model, as a function of the net charge $n-1$ of the composite, with $t'=0$ and the hole hopping matrix element $-t_h$, $t_h>0$. For each $n$, the composite with the lowest such energy is shown. The insets show the configuration of different composites. We use the darkness of the outline of a hole orbital to illustrate the probability of the hole. The hole is trapped on the orbitals that maximize electron-hole attraction.
	In both models, we find the energy is convex at $n=3$, which means the rollable 3e1h composite (red) lies below the dashed line through the 2e1h and 4e1h composites. This means the 3e1h composite will be stabilized against a 1:1 phase separation of 2e1h and 4e1h when the electron and hole densities are tuned to a ratio of 3:1. }
\label{fig:hole_kinetic}
\end{figure}

Indeed, there is a better way: we find that part of the kinetic energy can be used to stabilize the 3e1h composite. 
To see this point, let us consider what happens when switching on a hole hopping $t_h$, which is weak compared to the interaction strength. In either the triangle-kagome or AA-stacked triangle model, $t_h$ enables the hole to hop locally within the ``trap'' potential created by the three nearby electrons through the electron-hole attraction $V$. For example, for the 3e1h configuration given in Fig.~\ref{fig:hardest_competitors}(e), the hole is trapped within the three hole sites on top of the three electrons so that it optimizes the electron-hole attraction energy.
In this way, $t_h$ lowers the energy of the rollable 3e1h composite by the kinetic energy of the hole, and also lowers the energies of other types of composites, like 2e1h [Figs.~\ref{fig:hardest_competitors}(a) and (d)] or 4e1h [Figs.~\ref{fig:hardest_competitors}(c) and (f)], in a similar way. However, interestingly, in both the triangle-kagome and AA-stacked triangle models, we find that this kinetic energy gain from the hole's hopping happens to favor a three-electron-one-hole composite. This can be seen from a direct energetic calculation. We work out the kinetic energy gain of the $n$e1h composite ($n=2,3,4,\ldots$) in Fig.~\ref{fig:hole_kinetic} for both models, where one finds that the energy as a function of net charge is convex at 3e1h. This means that, when the densities of electrons and holes are tuned to 3:1, which is the most favorable situation for forming 3e1h composites, the rollable 3e1h composite has a lower kinetic energy than its nearby competitors, such as a 1:1 coexistence of 2e1h and 4e1h. As a result, at least in these two models, the hole's kinetic energy tends to select the 3e1h composite from among these competitors, which is what we previously relied on $U$ to achieve. In this sense, $t_h$ does help stabilize the rollable composite.

\begin{figure}[!tb]
\centering
\includegraphics[width=\columnwidth]{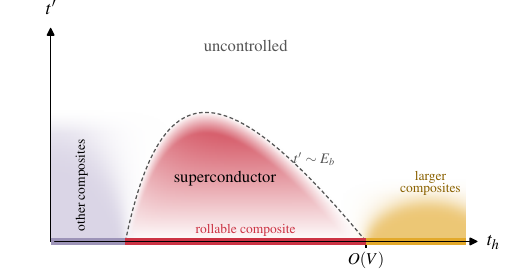}
\caption{Sketch of the phase diagram expected for both models in the plane of the hole hopping $t_h$ and the electron hopping $t'$. At $t'=0$ the composite transitions sharply from some configuration favored by interaction to the rollable 3e1h composite, and then to larger composites that are in general not rollable. At finite $t'$ these configurations usually cross over smoothly to each other. Superconductivity is expected to occur inside the red regime, with a $T_c$ that grows linearly with $t'$ at small $t'$, and to end somewhere as $t'$ approaches the scale of the third electron's binding energy, which is $O(t_h)$ in the triangle-kagome and AA-stacked triangle models. 
}
\label{fig:stability_sketch}
\end{figure}

Based on these analyses, we can now predict the phase diagram, which we sketch in Fig.~\ref{fig:stability_sketch} for both models. 
We first look at the line $t'=0$: 
At small $t_h$, interaction alone does not generally select the rollable composite. So generally the rollable composite occurs only when $t_h$ exceeds a threshold value, below which other unrollable composites are stabilized. 
If in some special settings the interactions alone already select the composite, there will be no lower threshold for the rollable composite. 
Upon increasing $t_h$, the rollable composite is destabilized when $t_h$ becomes comparable to the attraction $V$ that traps the hole. Further increasing $t_h$ makes the composite larger (yellow regime). We note that on the line $t'=0$ the electrons do not move, so all electrons sit on definite sites in the ground state. As a result, all transitions from one electron placement to another are sharp. Next, we consider what happens when $t'$ is switched on. With a finite $t'$, the different electron placements are generally connected by the Hamiltonian, so distinct composite configurations at $t'=0$ usually cross over smoothly to each other at finite $t'$.  Adding $t'$ on top of the rollable composite brings the system into the superconducting phase we are looking for (red shading). $T_c$ first increases with $t'$, since it is governed by the stiffness, which is proportional to $t'$ when $t'\ll t_h$. However, $t'$ also tends to unbind the electrons. When $t'$ approaches the binding energy of the third electron $E_b$, which is governed by $t_h$ in this scenario and decreases to 0 at large $t_h$, where larger composites become close in energy, the system smoothly enters an intermediate regime where the many-body wavefunction becomes a ``soup'' of 3e1h composites, 2e1h composites and single electrons. As a result, the superconductivity scenario which is based on the picture that electrons spend most of their time in rollable 3e1h composites no longer holds, so superconductivity is likely to get suppressed there.

\section{\texorpdfstring{Estimating $T_c$ by mapping to an effective boson model}{Estimating Tc by mapping to an effective boson model}}
\label{sec:boson_tc}

Next, we move on to estimate the $T_c$ inside the predicted superconducting phase. $T_c$ is determined by the superfluid stiffness, which is given by the density over the mass. 
 Therefore, below we will first map the electron-hole model onto a model of bosons, one boson for each rollable composite, find the mass $m_B$ of a boson and then study what the maximally allowed boson density is.

\begin{figure}[!tb]
\centering
\includegraphics[width=\columnwidth]{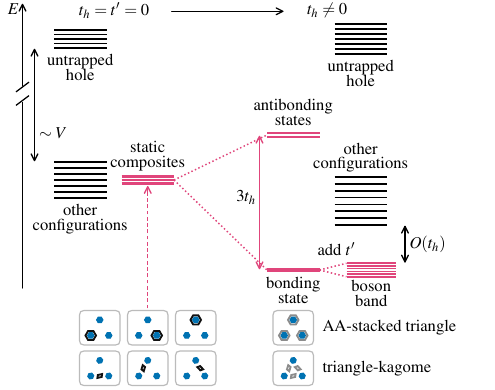}
\caption{Sketch of the energy levels of one composite. At $t_h=t'=0$, the degenerate static composites (pink) lie among the other configurations (black), and the states with an untrapped hole lie higher by about the attraction $V$ that traps the hole. Adding hole hopping $t_h$ splits the static composites into the bonding state and the antibonding states. The bonding state lies below all other states by a gap of order $t_h$. Switching on $t'$ then generates a band of width proportional to $t'$. The insets at the bottom illustrate the static composites and the bonding state of the AA-stacked triangle and triangle-kagome models.}
\label{fig:levels}
\end{figure}

The mapping from the electron-hole model to the boson model is subtle, because it needs a lowest-energy manifold that is isolated from all other states by a gap larger than the hopping of the bosons. Fig.~\ref{fig:levels} sketches how this manifold arises. At $t_h=0$, the static composites are degenerate. They may already be the lowest-energy manifold, if the interactions alone stabilize the composite, or they may not, if other configurations lie as low (Sec.~\ref{sec:interaction_stabilization}). Either way, $t_h$ splits the static composites into a bonding state and antibonding states. 
When $t_h$ is sufficiently large, the bonding state becomes the lowest-energy manifold, isolated from all other states by a gap of order $t_h$. We therefore project onto the bonding state. A boson is the composite in its bonding state, $t_h$ acts inside the boson, and only $t'$ moves it, which holds for $t'\ll t_h$. In the AA-stacked triangle and triangle-kagome models, for example, the three degenerate static composites on a triangle are split into a bonding state with energy $-2t_h$ and two degenerate antibonding states with energy $t_h$.

\begin{figure}[!tb]
\centering
\includegraphics[width=\columnwidth]{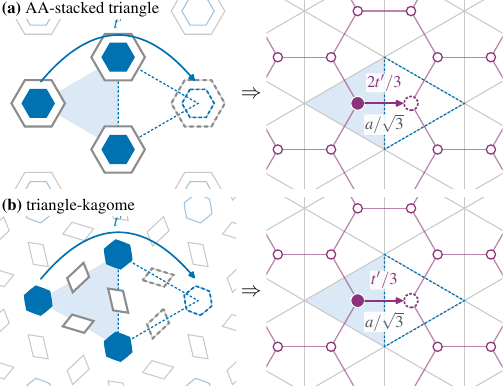}
\caption{Mapping from the original electron-hole model to the dual boson model, for (a) the AA-stacked triangle model and (b) the triangle-kagome model. Left: a composite in its bonding state, with its hole spread over its three orbitals (thick gray), and the electron hop $t'$ that carries it onto the neighboring triangle (dashed). Right: the corresponding hop of the boson on the dual honeycomb lattice of the centers of the triangular plaquettes. Its amplitude is $t'$ times the overlap of the two bonding states, $2/3$ in AA-stacked triangle and $1/3$ in triangle-kagome.}
\label{fig:boson_mapping}
\end{figure}

Mapping each bonding state to a boson site, we arrive at an effective boson model shown in
Fig.~\ref{fig:boson_mapping}. A boson occupies an elementary triangle of electrons, so we define the centers of these triangles as the boson sites, which form the dual honeycomb lattice of spacing $a/\sqrt3$. The electron's second-nearest-neighbor hop $t'$
	carries the composite onto a neighboring triangle, so it maps to a nearest-neighbor hop of the boson on its honeycomb lattice. Its amplitude $t_B$ is the matrix element of $t'$ between the bonding states on the two triangles, which is $2t'/3$ in the AA-stacked triangle model and $t'/3$ in the triangle-kagome model.
	From the band of the honeycomb lattice, the mass at the band bottom is $m_B=2\hbar^2/(t_Ba^2)$, that is,
\begin{align}
&m_B=\frac{3\hbar^2}{t'a^2}=27m_e\ \ \text{(AA-stacked triangle)},\label{eq:boson_mass_AA}\\
&m_B=\frac{6\hbar^2}{t'a^2}=54m_e\ \ \text{(triangle-kagome)},
\label{eq:boson_mass_bonding}
\end{align}
with $m_e=\hbar^2/(9t'a^2)$ the band mass of a free electron, which hops by $t'$ to its six neighbors at the distance $\sqrt3a$.

These big factors 27 and 54 should be understood as follows. Tracing the derivation back, $m_B/m_e=9\times2\times t'/t_B$. The factor 9 arises because the hop moves only one of the three electrons, so the boson advances by $a/\sqrt3$, a third of the electron's step $\sqrt3a$, and the mass scales as the inverse square of the step. The factor 2 arises because the boson has three nearest neighbors on its honeycomb lattice, whereas a free electron has six. The factor $t'/t_B$, $3/2$ in the AA-stacked triangle model and $3$ in the triangle-kagome model, is the inverse of the overlap of the two bonding states.
Therefore, these large factors of mass (27 and 54) are not completely a price of rolling: only the factor $t'/t_B$ is a price of rolling, since the factors 9 and 2 arise also for an unrollable composite, such as that of Fig.~\ref{fig:intro_rolling}(a), and are unavoidable for any composite of three electrons on a triangle. Rolling costs only this factor of order one, whereas the unrollable composite pays a factor of order the binding energy over the hopping.

We note that the mass estimated above is an asymptotic result in the small-$t_h$ and small-$t'$ limit. Increasing $t_h$ and $t'$ usually makes the composite lighter. With a finite $t_h$, the hole spreads somewhat beyond its three orbitals, which raises the overlap of the two nearby bonding states and tends to lower the mass. Upon increasing $t'$, the electrons also leave their triangle part of the time, and the composite moves more and more like its three electrons moving on their own, whose masses add up to $3m_e$.

Next, to estimate $T_c$ we need to study how high the density of the bosons can be pushed.
	The density of the bosons is bounded since bosons repelling each other tend to crystallize at high density. For bosons that repel as hard disks of diameter $\sigma$, crystallization happens at $n_B\sigma^2\simeq0.3$~\cite{xing_1990}. Thus, to know the maximal density of the bosons, we need to estimate the effective diameter of the bosons, which is their interaction range.

\begin{figure}[!tb]
\centering
\includegraphics[width=\columnwidth]{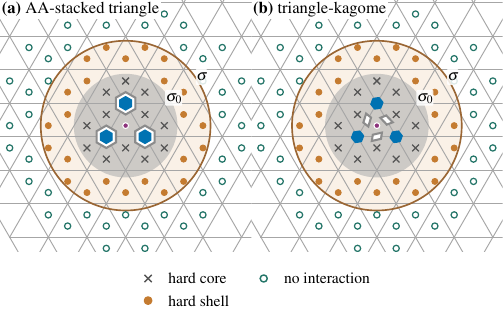}
\caption{Effective interaction range of bosons in (a) the AA-stacked triangle model and (b) the triangle-kagome model. We place one composite at the center, with its electrons filled and its hole orbitals outlined in gray. The purple dot represents this boson's position. 
When the second boson sits on boson sites marked by crosses, their electron and hole would overlap, which is not allowed, so the effective boson-boson repulsion is infinite. Therefore each boson has a hard core of radius $\sigma_0$ (gray shaded disk). When the second boson is placed on the orange sites, electrons of the two composites repel through nearest-neighbor repulsion, or their holes are within the range of the hole-hole repulsion. In both models, some of these repulsions are needed to keep each composite exactly 3e1h, and they are much stronger than the boson hopping. This gives each boson an impenetrable shell of radius $\sigma$, which is $\sim2.2a$ for both models.
Here we focus on the best-case scenario where those further-neighbor repulsions that are unnecessary for stabilizing the 3e1h composite are all screened, so there is no interaction outside the range of $\sigma$.
}
\label{fig:boson_repulsion}
\end{figure}

	Therefore, we study how two composites interact. As shown in Fig.~\ref{fig:boson_repulsion}, we find that a boson has a hard core in which the repulsion is infinite, and a hard shell, in which the repulsion is finite yet still impenetrable, as it is much stronger than the kinetic energy of the bosons. The hard core with a radius of $\sigma_0\approx1.3a$ exists because two nearby composites cannot share an orbital.
	 The hard shell with a radius $\sigma$ exists because of the electron-electron and hole-hole repulsions that are necessary to keep the composite exactly 3e1h.
	In the best-case scenario, where all unnecessary longer-range repulsions are screened,
	we find
	 $\sigma\approx2.2a$ in both models.

With these preparations, we now estimate the optimal $T_c$.
	$T_c$ is given by the BKT transition temperature, which is $T_c=\pi\hbar^2n_s/(2m_B)$, with $n_s$ the superfluid density just below $T_c$~\cite{nelson_kosterlitz_1977}. 
	Using the Monte Carlo results of Ref.~\cite{pilati_giorgini_prokofev_2008}, extrapolating from their Fig.~5 to $n_B\sigma^2\simeq0.3$, we get a maximal superfluid density $n_s^*\simeq0.8n_B$ just below $T_c$. As a result, the optimal $T_c$ is given by
\begin{equation}
T_c^*\simeq \frac{0.8 \pi \hbar^2 n_B}{2m_B}\simeq \frac{0.38\hbar^2}{m_B\sigma^2}.
\label{eq:tc_star}
\end{equation}
With the masses of Eqs.~\eqref{eq:boson_mass_AA} and \eqref{eq:boson_mass_bonding} and these values of $\sigma$, $T_c^*$ is about $3\%$ of $t'$ in the AA-stacked triangle model and about $1\%$ in the triangle-kagome model.

	The predicted $T_c$ grows in proportion to $t'$. 
But $t'$ cannot be too large as $t'$ also tends to pull the composite apart. 
When $t'$ approaches the scale of $t_h$ so that the kinetic energy gain of a free electron $6t'$ becomes comparable to the electron's binding energy inside the bound state, which is $O(t_h)$, the electron will have an order-one probability to run away. Then the scenario of superconductivity loses control as it is based on the picture of a tightly bound composite, and superconductivity is expected to terminate there.
We therefore expect $T_c$ to be maximized just below the upper edge of the superconducting phase, where $6t'$ approaches the binding energy of the third electron. We note in passing that in realistic systems, there is also a nearest-neighbor hopping $t$ that competes with binding $t_h$. It does not move the composite without deforming it, but, like $t'$, it lets an electron gain kinetic energy by leaving the composite. A nearest-neighbor hopping therefore leaves less room for $t'$, and the composite stays bound only if $t$ is also small compared with $t_h$.

\section{Realization and numerical evidence}
\label{sec:realization_gates}

We now ask how to realize the minimal models and the predicted superconductivity. An ideal setting should have two features. First, we need screening of the Coulomb interaction. This is not only because our analysis of the minimal models assumed short-range interactions, but also because a long interaction range would make the composites crystallize at a very low density and this limits $T_c$. The screening can come from a proximate metallic plane or from the Thomas-Fermi screening by the carriers of another conduction band. Second, we need a tunable noninteracting hole gap $\Delta$. As a reminder, our whole analysis of the stability of the 3e1h composite concerns the density ratio 3:1 of electrons and holes, the best case for the composite, and it assumes that in reality we have a way to tune $\Delta$ to reach this ratio.

One realistic setting that simultaneously satisfies the two requirements above
	is the dual-gate bilayer of Fig.~\ref{fig:single_gate_geometry}, widely used for two-dimensional materials such as moir\'e or non-moir\'e graphene multilayers and transition-metal dichalcogenides~\cite{zhou_2021_rtg,zhou_2021_rtg_sc,han_2026_wigner,xia_2025_twse2,guo_2025_twse2,gu2022dipolar,zhang2022correlated}. 
	In this setting, the electrons and the holes sit in two layers, which enables tuning the hole gap by an applied perpendicular electric field that is controlled by the voltage difference between the two gates. In the meantime, the gates screen the Coulomb interaction down to a distance comparable to the gate-to-sample distance.

We choose parameters to be the typical ones for moir\'e transition-metal dichalcogenides. The lattice constant is $a=10$~nm. The electron and hole layers are $d=0.1a$ apart, in a slab with the permittivities $\epsilon_\parallel=16$ in the plane and $\epsilon_z=8$ out of it, as in multilayer transition-metal dichalcogenides. The slab is encapsulated in hexagonal boron nitride with $\epsilon_1=4$~\cite{laturia_2018}, and the gates are $5$~nm away from the layers, $h_1=h_2=0.5a$. We define an energy scale $E_0=e^2/(4\pi\epsilon_0\epsilon_1a)\simeq36$~meV, the bare Coulomb interaction energy of two nearest-neighbor electrons, and measure all other energies in units of $E_0$. The form of interaction in such a setting resembles what we proposed in the minimal model: With the interaction derived in Appendix~\ref{app:gate_interactions}, when applied to the AA-stacked triangle model, we found $V_0\simeq4.9E_0$, which is much greater than the subleading interactions: $U_1 \simeq V_1\simeq W_1\simeq0.20E_0$. We checked that all these interactions are well screened, as they decay by two orders of magnitude up to a distance of $3a$.

\begin{figure*}[!tb]%
\centering
\includegraphics[width=\textwidth]{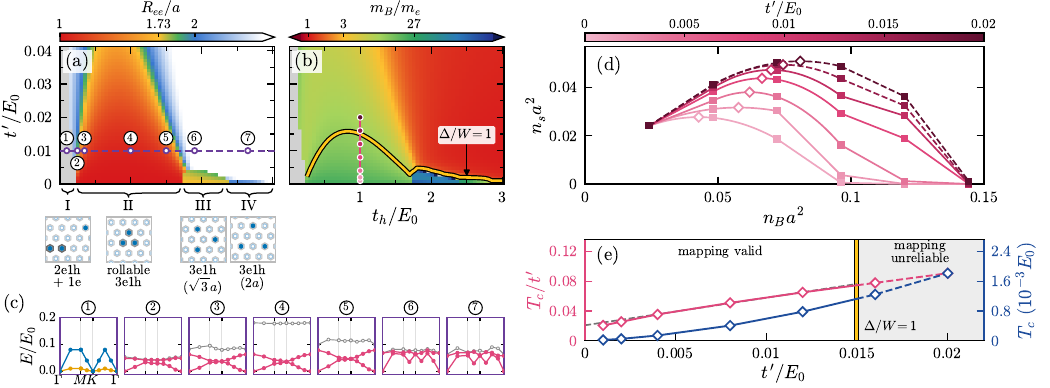}%
\caption{The AA-stacked triangle model in the dual-gate bilayer with $d=0.1a$, $\epsilon_\parallel=16$, $\epsilon_z=8$, $\epsilon_1=4$ and the gates at $h_1=h_2=0.5a$, from exact diagonalization on the torus of 48 electron sites. (a) The root-mean-square distance $R_{ee}$ of the electron pairs in the lowest state of three electrons and one hole. The numbers I to IV below the x axis mark four regimes with distinct ground-state configurations at $t'=0$. The gray area at low $t_h$ is where the ground state at an electron-hole filling ratio of 3:1 is 2e1h + 1e.
The panels below show the configuration at $t'=0$, where the darkness of the outline of a hole orbital represents the probability of the hole. 
(b) The mass $m_B$ of the composite in units of the band mass $m_e$ of a free electron, extracted from the curvature of the lowest band at $k=0$.  The yellow line marks where the direct gap $\Delta$ above the two bands of the composite equals their width $W$. Below the yellow line, $\Delta>W$, so the third band is well isolated from the two lowest bands that represent the overall motion of the composite bosons. This is the region where the mapping onto the boson model of Eq.~\eqref{eq:boson_model} is valid and is therefore the regime where our prediction of superconductivity is reliable. 
(c) The evolution of the excitation spectrum along the dashed purple line in (a). Panels (d), (e): We scan along the pink line in (b), at the points marked on it, to map out the superfluid density of the composite bosons and the optimal $T_c$. (d) The superfluid density $n_s$ as a function of the boson density $n_B$. Each curve is obtained for one sampling point on the pink line cut, marked by dots in (b). We obtain $n_s$ by exact diagonalization of the boson model on the 96-site honeycomb lattice at each given boson number. The maximal $n_s$ for each $t'$ is marked by a diamond on each curve.
The curves are dashed outside the controlled region. (e) The optimal $T_c$ in units of $t'$ (left axis) and of $E_0$ (right axis), obtained from the largest $n_s$ in (d) through $T_c=\pi\hbar^2n_s/(2m_B)$. The gray region beyond the yellow line is outside the controlled region. The gray dashed line is a linear fit of $T_c/t'$, extrapolated to $t'=0$.
}%
\label{fig:tc_plane}
\end{figure*}

Our remaining goal is to show that what we estimated in the minimal model is indeed realizable in this realistic setting. For that, we now use exact diagonalization (ED) to study the phase diagram of this realistic model.

The main results of exact diagonalization are summarized in Fig.~\ref{fig:tc_plane}, which are obtained on the torus of 48 electron sites. To see in what regime of $t_h$ and $t'$ the composite can be stabilized in the ground state, we map out the averaged electron-electron distance in the ground state, which is shown in Fig.~\ref{fig:tc_plane}(a), which is a direct way to show how the ground-state configuration evolves. %
We first look at the evolution of the ground state along the axis $t'=0$. We find, on this axis, the system starts by favoring a phase separation of 2e1h and a single electron at low $t_h$ (I). Upon increasing $t_h$, the third electron is bound into the composite, forming the rollable composite (II) so $R_{ee}\approx a$. Further increasing $t_h$ makes the ground state abruptly transition into expanded composites at larger $t_h$ (III and IV) as indicated by the $R_{ee}$ value, which is approximately $\sqrt{3}a$ and $2a$ in the two regimes. 
Thus, the whole evolution along $t'=0$ is in agreement with our prediction.

Switching on $t'$, we find 
the distinct static-electron configurations in regimes II, III and IV melt and cross over to each other continuously, as shown by the averaged electron-electron distance $R_{ee}$ in Fig.~\ref{fig:tc_plane}(a).\footnote{Here, the transitions into III remain sharp, but this is an artifact of the minimal model. %
	Namely, as we do not have a nearest-neighbor hopping of the electrons in the model, the configurations in III and those in II and IV are not connected by the Hamiltonian. In reality, with a nearest-neighbor hopping of the electrons, the transitions are always a smooth crossover.} %
	In this figure, $t'$ spreads these regimes into dome-shaped areas. The rollable composite in regime II is bound strongly by $t_h$, so it has the highest dome, whereas the binding of the larger composites (III and IV) is much weaker, so their domes are much smaller.

	It is instructive to look into the evolution of the excitation spectrum. We map out the evolution of the spectrum along a line cut at $t'=0.01E_0$, as shown in Fig.~\ref{fig:tc_plane}(c). At low $t_h$ (\cnum{1}), the dispersion consists of contributions from a single electron (blue) and the 2e1h composite (orange). %
	Upon increasing $t_h$ (\cnum{2} to \cnum{4}), the system enters regime II, where a gap opens above the two lowest bands (pink), while the dispersion of these two bands becomes similar to the honeycomb tight-binding dispersion, as suggested by the Dirac crossing at $K$. This is a direct manifestation of the effective boson representation we introduced in Sec.~\ref{sec:boson_tc} [Fig.~\ref{fig:boson_mapping}]. %
	Further increasing $t_h$ (\cnum{5} to \cnum{7}) closes the gap again, but this time, the band evolves into an electron-like band (resembling the blue line at \cnum{1}), which is again in agreement with the picture: the binding of the electrons is loosened when the composite expands, and it eventually breaks into individual electrons.

	Next, we map out the mass $m_B$ of the composite.
	The boson mass $m_B$ can be extracted from the curvature of the excitation spectrum near $k=0$: when the 3e1h composite is bound, the curvature of the lowest-lying band gives the boson mass, as this band describes the overall motion of the composite.
	We show the extracted mass in Fig.~\ref{fig:tc_plane}(b). %
	In regime II we see that it approaches about $25m_e$ as $t'\to0$ at the left end of II, close to the $27m_e$ that Eq.~\eqref{eq:boson_mass_AA} predicts for $t_h\to0$. %
	In comparison, for larger composites (see regimes III and IV) the ratio $m_B/m_e$ diverges as $t'\to0$, which can be attributed to the absence of rollability. Finite $t_h$ and $t'$ tend to make $m_B/m_e$ smaller, as it needs to cross over to the single-electron mass at the upper right corner.

To check our prediction of $T_c$, we next extract numerically the superfluid density $n_s$ of the ground state, and use the Berezinskii--Kosterlitz--Thouless (BKT) relation $T_c=\pi\hbar^2n_s/(2m_B)$~\cite{nelson_kosterlitz_1977} as an estimate for $T_c$. For each $(t_h,t')$, we are interested in the optimal $T_c$ achievable, so we need to find the carrier density that maximizes the superfluid density, which requires scanning the number of carriers up to the crystallization threshold. However, this is numerically infeasible if we proceed directly in the original fermion model, as the Hilbert space is exponentially large at such high particle numbers. We can still do it, though, by first mapping onto a boson model and then solving the boson model by exact diagonalization. In this way, four particles in the original model become one particle in the boson model, which greatly reduces the dimension of the Hilbert space and makes the density regime of interest accessible.

But how do we do the fermion-to-boson mapping numerically? We first solve the momentum-resolved excitation spectrum by exact diagonalization, and then use the wavefunctions of the lowest two ``many-body bands'' (the pink bands in Fig.~\ref{fig:tc_plane}(c)), which are essentially the boson bands, to construct two Wannier orbitals, one on the triangles pointing up and one on those pointing down. These Wannier functions can be interpreted as the wavefunctions of local bosons, which live on the centers of the triangles, the honeycomb lattice dual to the electron lattice (Fig.~\ref{fig:boson_mapping}). For the torus of 48 electron sites that we use, this is the honeycomb lattice of 96 sites. We then project the Hamiltonian onto these Wannier orbitals, and arrive at an extended Hubbard boson model
\begin{equation}
H_B=-\sum_{i\neq j}t_{ij}b_i^\dagger b_j+\sum_{i<j}U_{ij}n_in_j ,
\label{eq:boson_model}
\end{equation}
where $t_{ij}$ and $U_{ij}$ are the hopping and the interaction of the bosons. Namely, we calculate the boson hopping as $t_{ij}=-\langle W_i|H|W_j\rangle$, where $|W_i\rangle$ is the Wannier orbital state,
\begin{equation}
|W_i\rangle=b_i^\dagger|0\rangle=\sum_{i_1<i_2<i_3,i'}W_i(i_1,i_2,i_3;i')c_{i_1}^\dagger c_{i_2}^\dagger c_{i_3}^\dagger h_{i'}^\dagger|0\rangle ,
\label{eq:composite_state}
\end{equation}
where $i_1,i_2,i_3$ are the sites of the three electrons, $i'$ is the site of the hole, and $W_i(i_1,i_2,i_3;i')$ is the wavefunction of the composite that we obtained from the Wannier construction of the many-body bands described above. The boson interaction is given by $U_{ij}=\langle W_iW_j|H|W_iW_j\rangle-\langle W_i|H|W_i\rangle-\langle W_j|H|W_j\rangle$, where $|W_iW_j\rangle$ is the state with one boson in $W_i$ and one in $W_j$. Here we note that the two-boson state $|W_iW_j\rangle$ is obtained by normalizing the naive two-body state $b_i^\dagger b_j^\dagger|0\rangle$. The justification and the idea underlying this procedure are detailed in Appendix~\ref{app:boson_ed}.

With this boson model, we then obtain the ground-state energy by another round of exact diagonalization at a fixed number of bosons, and extract $n_s$ from the energy of the ground state under a twist $\theta$ of the boundary,
\begin{equation}
E(\theta)-E(0)=\frac{S}{2}\frac{\hbar^2n_s}{m_B}\left(\frac{\theta}{L}\right)^2 ,
\label{eq:twist}
\end{equation}
with $S$ the area of the system and $L$ its width along the twist.%

We note that the whole fermion-boson mapping procedure described above is valid only where the two bands are well isolated from the states above them. We take this to be the case where the direct band gap $\Delta$ between the second band and the third band is at least the width $W$ of the two bands, which is the regime below the yellow line in Fig.~\ref{fig:tc_plane}(b). We consider our fermion-boson mapping results valid only in this regime. 

We carry out this procedure for a few points sampled along a line cut through the predicted regime of superconductivity, at $t_h=E_0$, the pink line in Fig.~\ref{fig:tc_plane}(b). For each sampled value of $t'$, we scan the number of bosons and extract $n_s$, as shown in Fig.~\ref{fig:tc_plane}(d). We see that the superfluid-to-crystal transition typically occurs around $n_Ba^2\simeq0.14$. Above it, the bosons form a crystal, i.e., a charge-density wave. The crystal phase features no stiffness under such a twist because, in a crystal, the twist enters only through moving the whole crystal by one period, so that $E(\theta)-E(0)\propto e^{-L/\xi}(1-\cos\theta)$ vanishes exponentially as the torus grows, where $\xi$ is the correlation length of the boson field. Below this threshold, we find that $n_s$ can get up to $\sim0.05/a^2$. We then map out $T_c$ accordingly in Fig.~\ref{fig:tc_plane}(e), where we find that $T_c$ is a few percent of $t'$, in agreement with the estimate of Sec.~\ref{sec:boson_tc}. Here we see that $T_c/t'$ increases with $t'$, which is mainly due to the decrease of the mass upon increasing $t'$, as shown in Fig.~\ref{fig:tc_plane}(b). This is reasonable, because a larger $t'$ increases the amplitude of higher-order processes, in which the electrons leave the triangle and the composite moves more like its electrons on their own. Notably, we find $T_c/t'$ stays a few percent when extrapolated to the limit $t'=0$, rather than vanishing%
\footnote{
We note that, at small $t'$, $T_c/t'$ has a weak tendency to bend downward. This correction arises from the exponential tail of the interaction between the bosons, $U_{ij}\propto e^{-|\mathbf r_i-\mathbf r_j|/\lambda}$, which strictly speaking takes $T_c/t'$ down to zero as $t'\to0$, though only by a logarithm. However, this 
correction is small: it shows up only below $t'_*=E_0e^{-2a/\lambda}$, which is exponentially low for a short decay length $\lambda$, and in Fig.~\ref{fig:tc_plane}(e) only at the smallest $t'$.  This point is detailed in Appendix~\ref{app:tail}.
}%
	, as shown by the linear fit of $T_c/t'$ (gray dashed line in Fig.~\ref{fig:tc_plane}(e)). This is exactly what we predicted in theory. Since $T_c$ grows with $t'$, it is maximized at the largest $t'$ before the theory loses control. The controlled regime is marked by the yellow line in Fig.~\ref{fig:tc_plane}(e), under which we find an optimal $T_c\simeq1.1\times10^{-3}E_0$, which is about $0.5$~K for a van der Waals system with $a=10$~nm. Beyond the yellow line, the theory loses control as the mapping to bosons is no longer reliable, so we cannot safely tell whether $T_c$ will keep rising further or get suppressed there.

It is possible to achieve an even higher $T_c$ in reality. What one can do is to realize it in a natural crystal setting with lattice scale $a\sim3$~\AA\ rather than the typical 2D van der Waals setting we considered above, where $a\sim10$~nm. In a natural crystal, 
for the same dielectric constants, $E_0\propto1/a$ becomes about $30$ times larger. Provided that $t_h$ and $t'$ scale with $E_0$, the optimal $T_c$ rises by the same factor, to about $15$~K. The dielectric constants can also be smaller than the values we used above. In that case, the $T_c$ can be pushed to the order of a hundred kelvin in optimal cases. %
	However, the price is that we can no longer rely on gates to screen the interactions, as the typical gate-to-sample distance cannot be pushed below the nm scale. Yet, the screening can still be achieved by the Thomas-Fermi screening from the carriers of another band inside the material itself.
As for the hole gap, it can still be tuned by a perpendicular electric field, as in bilayer graphene%
, but only by a few tens to a hundred meV, a fraction of $E_0$.
There are alternative ways to achieve a tunable gap, such as applying pressure normal to the plane, which can change the distance of atoms in nearby atomic layers, and thus affect the on-site potential of electron and hole orbitals.

\section{Conclusion}
\label{sec:conclusion}

In summary, we have shown that, when the lattice symmetry enables the rolling motion of a charge-$2e$ composite, the composite's effective hopping amplitude is set by the single-particle hoppings, so $T_c$ no longer vanishes in the strong-binding limit but approaches a fraction of the electron bandwidth [Fig.~\ref{fig:intro_rolling}(e)].
Whether a composite can roll is determined by symmetry alone. As a result, all minimal lattice models that host rollable 3e1h composites can be constructed once and for all [Fig.~\ref{fig:hopping_models}], and they serve as prototypes in the search for such superconductors. Whether the rollable composite can actually form is a question of energetics, whose answer depends on the model. 
We have shown that the kinetic energy of the hole can be used to stabilize these composites in some models. We propose to realize this scenario in
a dual-gate bilayer setting, and have verified the proposal by exact numerical results. These analyses show the potential of using rollability as a guiding principle for synthesizing Coulomb-driven superconductors with an appreciable $T_c$.

One interesting direction is to consider a 3D rollability problem. The proof of the criterion in Sec.~\ref{sec:theory} carries over directly, with planes added to points and lines: a composite pattern $C_0$ is 3D-rollable on a lattice $L$ if and only if no point is left in place by all elements of $\mathcal G(C_0,L)$ and no line or plane is mapped onto itself by all of them. But one still needs to construct the minimal lattice models in 3D.

\begin{acknowledgements} 
We thank Patrick A. Lee and Jason Alicea for discussion.
We used the AI tools Claude (Anthropic), ChatGPT (OpenAI), DeepSeek, and Kimi (Moonshot AI) to write the code for the exact diagonalization and for the figures, and to edit the text. We directed this work, checked all AI-generated content and results, and take full responsibility for them.
Z. D. acknowledges support from the Gordon and Betty Moore Foundation’s EPiQS Initiative, Grant GBMF8682.
\end{acknowledgements}

\newpage
\appendix

\section{Screened Coulomb interaction in dual-gate geometry}
\label{app:gate_interactions}
Here we give the screened Coulomb interaction in the geometry of Fig.~\ref{fig:single_gate_geometry}. The electron and hole layers are the top and bottom atomic layers of a slab of thickness $d$, with the permittivity $\epsilon_\parallel$ in the plane and $\epsilon_z$ out of the plane, or $\epsilon$ if the slab is isotropic. The dielectric $\epsilon_1$ fills the distances $h_1$ and $h_2$ between the slab and the top and bottom gates. For each in-plane wave number $q$, the potential in each region is a combination of $e^{\pm qz}$, and of $e^{\pm qz\sqrt{\epsilon_\parallel/\epsilon_z}}$ in the slab. Matching the potential and the normal electric displacement at the two layers, with zero potential at the gates, relates the potentials $\phi_e$ and $\phi_h$ of the two layers to their charge densities $\sigma_e$ and $\sigma_h$ through $\epsilon_0qK(q)\,(\phi_e,\phi_h)=(\sigma_e,\sigma_h)$, with the symmetric matrix
\begin{equation}
\begin{aligned}
K_{ee}&=\epsilon_1\coth(qh_1)+\sqrt{\epsilon_\parallel\epsilon_z}\coth\bigl(qd\sqrt{\epsilon_\parallel/\epsilon_z}\bigr),\\
K_{hh}&=\epsilon_1\coth(qh_2)+\sqrt{\epsilon_\parallel\epsilon_z}\coth\bigl(qd\sqrt{\epsilon_\parallel/\epsilon_z}\bigr),\\
K_{eh}&=-\sqrt{\epsilon_\parallel\epsilon_z}\Big/\sinh\bigl(qd\sqrt{\epsilon_\parallel/\epsilon_z}\bigr).
\end{aligned}
\label{eq:slab_matrix}
\end{equation}%
The screened Coulomb interactions between two electrons, between an electron and a hole, and between two holes, at in-plane distance $r$, are then
\begin{equation}
\begin{pmatrix}U(r)&V(r)\\V(r)&W(r)\end{pmatrix}
=\frac{e^2}{2\pi\epsilon_0}\int_0^\infty dq\,J_0(qr)\,K(q)^{-1},
\label{eq:slab_interactions}
\end{equation}
where $J_0$ is the Bessel function. With the dual gate, these interactions decay exponentially at distances large compared with the distance between the gates. We note parenthetically that the interaction of each charge with its own image only shifts the energies of single electrons and holes, and these shifts are absorbed into $\Delta$ and the chemical potential.

\section{Exact diagonalization of the AA-stacked triangle model in the dual-gate bilayer}
\label{app:ed}

	In this appendix, we explain how we carry out this exact diagonalization, and how we read the results of Fig.~\ref{fig:tc_plane}(a) and (b) from it.

We solve the model on the torus of 48 electron sites [Fig.~\ref{fig:ed_sectors}(a)], with the parameters of Fig.~\ref{fig:tc_plane} and the interactions of Appendix~\ref{app:gate_interactions} summed over their periodic images. At each point $(t_h,t')$, we solve a set of sectors $s=(n,m)$, each with a fixed number $n$ of electrons and $m$ of holes. 
	At $t'\neq0$, we can solve only sectors with few particles. Namely, the hop $t'$ moves an electron by $\sqrt3a$, so it keeps the electron on its sublattice of spacing $\sqrt3a$. The numbers of electrons on the three sublattices are therefore conserved. Together with the momentum, these numbers split each sector into blocks, which we diagonalize separately. These blocks grow rapidly with $n$ and $m$, which limits the sectors we solve at $t'\neq0$ to those marked in Fig.~\ref{fig:ed_sectors}(b). In comparison, at $t'=0$ we can solve far larger sectors. Namely, the electrons do not move at $t'=0$, so each arrangement of the electrons is an isolated block of its own, in which only the holes move. We can therefore solve systems of up to 192 electron sites along $t'=0$. 

\begin{figure*}[t]%
\centering
\includegraphics[width=\textwidth]{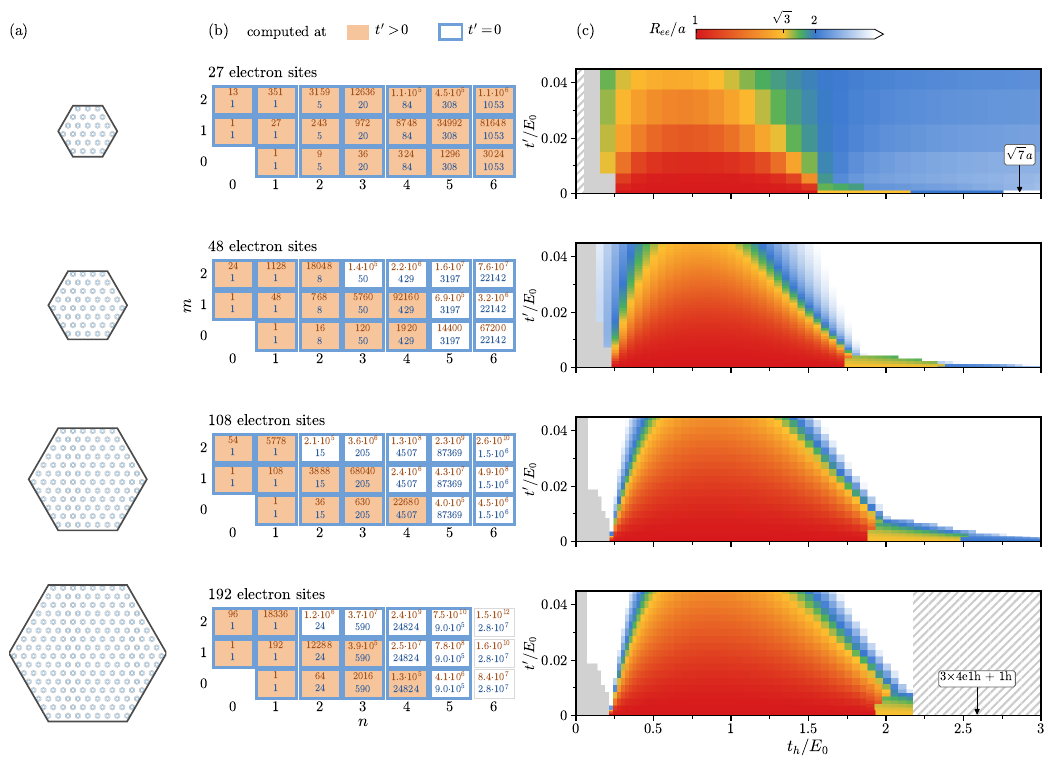}
\caption{Results of exact diagonalization of the model of Fig.~\ref{fig:tc_plane} on the tori of 27, 48, 108 and 192 electron sites. From left to right: (a)~the lattice, with periodic boundary conditions; (b)~the sectors of $n$ electrons and $m$ holes that we solve, at $t'=0$ for those framed in blue and at $t'\neq0$ for those filled in orange; (c)~the phase diagram at the ratio 3:1 in the plane of $t_h$ and $t'$, colored by $R_{ee}$ as in Fig.~\ref{fig:tc_plane}(a). In (b), the numbers in each cell are the size of the largest block at $t'\neq0$ (upper) and the number of arrangements of the electrons up to the symmetries of the torus (lower). In (c), the gray area is where the ground state is not in the 3e1h sector. The hatched area is where the ground state is not computed. Although the looser configurations at larger $t_h$ compete with one another, so that the favored one might depend on system size, the rollable 3e1h is stable: regime II only widens as the torus grows.}
\label{fig:ed_sectors}
\end{figure*}

\begin{figure*}[t]%
\centering
\includegraphics[width=\textwidth]{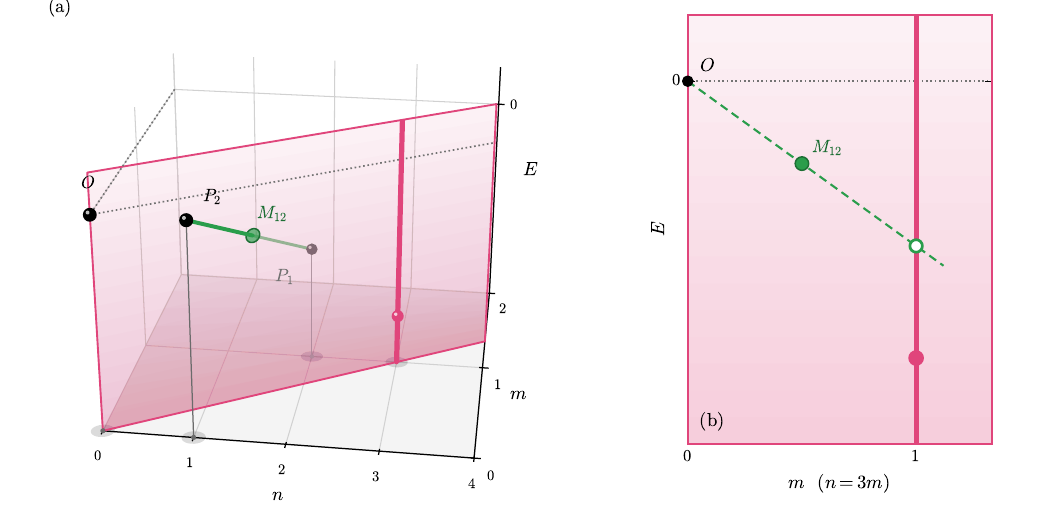}%
\caption{A schematic illustration of the method we used to compare the energies of states in sectors with different electron and hole numbers at a given electron-hole density ratio (3:1 in this case). At this given density ratio, the system can either be pure 3e1h, or a phase separation consisting of a mixture of two sectors, denoted as $s_1=(n_1,m_1)$ and $s_2=(n_2,m_2)$. To see which one wins, we need to first obtain the mixing ratio and then rescale the energy to 3e1h to compare. The first step is done in (a), where we use points $P_1=(n_1,m_1,E_1)$ and $P_2=(n_2,m_2,E_2)$ to denote the states of interest in the two sectors. The line connecting them shows how energy changes upon varying their mixing ratio. The intersection of this line with the vertical plane $n=3m$ (shaded), which is marked by a green dot and denoted by $M_{12}$, corresponds to the energy of their mixture at an e/h density ratio of 3:1.
However, this mixture state $M_{12}$ still does not have the same numbers of electrons and holes as 3e1h, so we cannot compare their energy directly. To compare directly, we need to first rescale $M_{12}$ to 3e1h. This is done in (b), where we focus on the shaded plane of (a), and rescale $M_{12}$ to 3e1h by connecting the origin $O$ with $M_{12}$ (dashed green line) and extending it to the vertical line at $m=1$, $n=3$. The height of the intersection (green circle) represents the energy of the mixture state. This energy can be directly compared with the energy of a state in the 3e1h sector (pink dot).}
\label{fig:schematic_3to1}
\end{figure*}

Next, we explain how to determine the ground state at the ratio 3:1 of electrons to holes, which gives the regimes I to IV and the gray points of Fig.~\ref{fig:tc_plane}. 
The ground state at this electron-hole density ratio does not have to be a single state with three electrons per hole. The system can also phase separate into a mixture of two states, such as a mixture of 2e1h composites and single electrons. We compare all these possibilities through a graphical method sketched in Fig.~\ref{fig:schematic_3to1}. The idea of this method is that the energy of such a phase separation interpolates between the energies of the two states, at the mixing ratio that gives three electrons per hole. Rescaled to three electrons and one hole, it compares directly with the single states at this ratio, and the lowest one is the ground state at 3:1.

One question that needs to be addressed before proceeding is whether we need to worry that the system might favor forming larger composites that our numerics is not able to study, given that we can study only a limited number of particles at general $t'$ (up to 4e1h). Below we show that it is not a problem. The reason is, first, that at $t'=0$ we can really study sectors of very large particle numbers (such as 6e2h),  where we indeed find 3e1h is the ground state in regime II [Fig.~\ref{fig:ed_sectors}(c)].
Meanwhile, we have a physical argument that so long as larger composites do not form at $t'=0$, they are unlikely to form at finite $t'$. This is because $t'$ naturally favors rollable composites as they gain $O(t')$ kinetic energy, whereas the larger composites are usually unrollable. 
We check this numerically on a smaller torus of 27 electron sites, where at finite $t'$ we can still solve the sectors up to 6e2h. As shown in the 27-site panel of Fig.~\ref{fig:ed_sectors}(c), the 3e1h ground state always wins over the 6e2h sector when $t'$ is switched on starting from regime II on the $t'=0$ axis. Therefore, it is safe to say that adding $t'$ does not select larger composites over the rollable 3e1h one.

Next, we explain how we measure the size $R_{ee}$ of the composite in Fig.~\ref{fig:tc_plane}(a). We measure it in the ground state $|\psi\rangle$ of the 3e1h sector. Since this state has zero momentum, the composite in it is spread evenly over the torus, so the electron density does not show how large the composite is. Therefore, we have to extract the root-mean-square distance between two electrons from the short-range correlation in the many-body wavefunction. Namely, we define $R_{ee}$ through
\begin{equation}
R_{ee}=\sqrt{\frac{1}{3}\langle\psi|\sum_{i<j}d_{ij}^2n_in_j|\psi\rangle},
\end{equation}
where $n_i=c_i^\dagger c_i$ is the number of electrons on site $i$, and $d_{ij}$ is the distance between the sites $i$ and $j$. On the torus, we take $d_{ij}$ as the shortest distance between the two sites.

\section{Mapping the fermion model onto the boson model}
\label{app:boson_ed}

As a reminder, in Sec.~\ref{sec:realization_gates} we map the original fermion model onto the extended Hubbard boson model
\begin{equation*}
H_B=-\sum_{i\neq j}t_{ij}b_i^\dagger b_j+\sum_{i<j}U_{ij}n_in_j ,
\end{equation*}
where $t_{ij}$ and $U_{ij}$ are the hopping and the interaction of the bosons. In this appendix, we explain how this mapping is done, i.e., how we obtain $t_{ij}$ and $U_{ij}$ from the original fermion model.

We first obtain the boson hopping $t_{ij}$. From the two bands of the composite computed in Appendix~\ref{app:ed} (the pink bands in Fig.~\ref{fig:tc_plane}(c)), we construct two Wannier orbitals, one on the triangles pointing up and one on those pointing down. To do so, we first select columns of the density matrix of the two bands~\cite{damle_lin_ying_2015}, and then minimize the spread of the Wannier orbitals~\cite{marzari_vanderbilt_1997,marzari_2012}. The centers of the Wannier orbitals form the honeycomb lattice of 96 sites, on which the bosons live. A boson in the Wannier orbital $W_i$ is a composite of three electrons and one hole, in the state
\begin{equation}
|W_i\rangle=b_i^\dagger|0\rangle=\sum_{i_1<i_2<i_3,i'}W_i(i_1,i_2,i_3;i')c_{i_1}^\dagger c_{i_2}^\dagger c_{i_3}^\dagger h_{i'}^\dagger|0\rangle ,
\label{eq:boson_operator}
\end{equation}
as in Eq.~\eqref{eq:composite_state}. Here $W_i(i_1,i_2,i_3;i')$ is the wavefunction of the Wannier orbital $W_i$, namely the amplitude for the three electrons to sit on the sites $i_1,i_2,i_3$ and for the hole to sit on the site $i'$. This wavefunction is normalized, $\sum|W_i(i_1,i_2,i_3;i')|^2=1$. The boson hopping $t_{ij}=-\langle W_i|H|W_j\rangle$ then follows from the two bands, by a Fourier transform into the basis of the Wannier orbitals. We keep every hopping down to $10^{-3}$ of the largest one, i.e., the nearest-neighbor hopping on the boson's honeycomb lattice. 

Next, we obtain the boson interaction $U_{ij}$, which can be calculated through $U_{ij}=\langle W_iW_j|H|W_iW_j\rangle-\langle W_i|H|W_i\rangle-\langle W_j|H|W_j\rangle$~\cite{koshino_2018,ciuti_1998,combescot_2008}. Therefore, we need to first obtain (or construct) the two-boson state $|W_iW_j\rangle$, in which one boson is in $W_i$ and the other is in $W_j$. Naively, we might think that the two-boson state is just $b_i^\dagger b_j^\dagger|0\rangle$, which by Eq.~\eqref{eq:boson_operator} reads
\begin{multline}
b_i^\dagger b_j^\dagger|0\rangle=\sum W_i(i_1,i_2,i_3;i')W_j(j_1,j_2,j_3;j')\\
\times c_{i_1}^\dagger c_{i_2}^\dagger c_{i_3}^\dagger h_{i'}^\dagger c_{j_1}^\dagger c_{j_2}^\dagger c_{j_3}^\dagger h_{j'}^\dagger|0\rangle ,
\label{eq:two_boson_naive}
\end{multline}
where $j_1,j_2,j_3$ are the sites of the electrons of boson $j$, $j'$ is the site of the hole of boson $j$, and the sum runs over the sites of both bosons. However, $b_i^\dagger b_j^\dagger|0\rangle$ is in general not normalized, which reflects a big problem. 
Before digging into it, let us first explain how we see it is not normalized: 
when the two bosons come close, some terms of Eq.~\eqref{eq:two_boson_naive} put an electron of boson $i$ and an electron of boson $j$ on the same site, $i_m=j_n$, or put the two holes on the same site, $i'=j'$. The Pauli principle removes all such terms, since $c_{i_m}^\dagger c_{i_m}^\dagger=0$. As a result, the norm of $b_i^\dagger b_j^\dagger|0\rangle$ is then no longer one.

How, then, do we obtain $|W_iW_j\rangle$? Can we just normalize $b_i^\dagger b_j^\dagger|0\rangle$ by hand, $|W_iW_j\rangle=b_i^\dagger b_j^\dagger|0\rangle/\|b_i^\dagger b_j^\dagger|0\rangle\|$? This is of course a quite artificial procedure. But it is justified in the sense that it has the same effect on the low-energy physics as the true physics does. 

To see this point, let us first look at two bosons far apart. For two such bosons, the norm of $b_i^\dagger b_j^\dagger|0\rangle$ is close to one, so the normalization changes almost nothing, which is good. Our only concern is therefore two bosons that come close. What happens to two such bosons in reality? Two composites that come close deform each other. Such configurations of deformed composites go beyond the boson model, since a boson is a composite in its bonding state (Sec.~\ref{sec:boson_tc}). However, these configurations are usually high in energy, because they excite the composites internally. Therefore, these configurations are irrelevant to the low-energy physics. This means that all the boson model has to do is to keep the bosons out of these configurations.
The normalization procedure described above indeed keeps two bosons from coming close. For two bosons that come close, the Pauli principle has removed the terms of Eq.~\eqref{eq:two_boson_naive} in which the two composites overlap. The normalization then brings the total weight of the remaining terms back to one. The weight is thus ``squeezed'' into the remaining configurations at the edge of the overlap, where the particles inside two composites still sit close to each other. In these configurations, the two composites repel each other strongly, so $U_{ij}$ usually comes out large. Therefore, the $U_{ij}$ obtained by our normalization procedure also keeps the bosons apart, just as they should behave in reality. We note that our argument relies on the assumption that two composites do not bind into a state of six electrons and two holes. This is indeed the case, as discussed in Appendix~\ref{app:ed}.

We evaluate $U_{ij}$ as described above, keeping in each Wannier orbital the configurations that carry $99.9\%$ of its weight. In the actual computation, to speed it up, we further take $U_{ij}$ to be infinite wherever it exceeds $E_0$, which is more than 35 times the largest hopping $|t_{ij}|$ everywhere along the pink line in Fig.~\ref{fig:tc_plane}(b). This removes the states in which two bosons sit on two sites $i$ and $j$ with infinite $U_{ij}$, and makes the Hilbert space dimension up to ten times smaller. 
Meanwhile, we suppress all $U_{ij}$ below $10^{-4}$ times the largest $|t_{ij}|$, which are too small to affect the bosons.

\section{\texorpdfstring{Logarithmic suppression of $T_c/t'$ at extremely low $t'$ due to the exponential tail of the interaction}{Logarithmic suppression of Tc/t' at extremely low t' due to the exponential tail of the interaction}}
\label{app:tail}

In a footnote of Sec.~\ref{sec:realization_gates} we note that $T_c/t'$ in Fig.~\ref{fig:tc_plane}(e) has a weak tendency to bend downward at small $t'$. Strictly speaking, this bend takes $T_c/t'$ down to zero as $t'\to0$, though only by a logarithm. In this appendix, we explain this point.

The logarithmic suppression comes from a difference between the realistic setting and the model with short-range interactions that we considered in theory (Sec.~\ref{sec:boson_tc}). Namely, in the realistic setting the interactions have an exponential tail, $U_{ij}\propto e^{-|\mathbf r_i-\mathbf r_j|/\lambda}$ with $\lambda\simeq0.4a$. %
 This tail has two origins. First, the interaction screened by the gates still has an exponential tail (Appendix~\ref{app:gate_interactions}). Second, at finite $t_h$ and $t'$, the wave function of a composite spreads beyond its triangle with an exponential tail. Two composites then also interact through the overlap of their wave functions, which decays exponentially as well. Our analysis of Sec.~\ref{sec:boson_tc} assumes small $t_h$ and $t'$ with every longer-range repulsion screened. In this limit, the tail is absent. However, a realistic setting cannot reach this limit, so some exponential tail is always present.

This tail of repulsion leads to a logarithmic suppression of $T_c/t'$ at extremely small $t'$. When the kinetic energy of the bosons ($\propto t'$) falls below the exponentially small potential, the potential becomes impenetrable. Therefore, the boson's effective impenetrable radius is $\sigma\simeq\max[2a,\lambda\ln(E_0/t')]$, so that by Eq.~\eqref{eq:tc_star} $T_c/t'$ asymptotically decreases as $a^2/[\lambda^2\ln^2(E_0/t')]$ at low $t'$. However, note that this logarithm is visible only when $t'$ is below the scale $t'_* = E_0 e^{-2a/\lambda}$, where $\lambda\ln(E_0/t')$ exceeds the radius of the short-range repulsion, $\sim 2a$. 
In the regime of interest, this scale $t'_*$ is usually quite low because (a) for the relevant $t'$ and $t_h$, the electrons and the hole are tightly bound, so that the wave function leakage to farther sites can only give a $\lambda\ll a$, and (b) the original electron-electron interaction is well screened, so it will also give a $\lambda\ll a$. As a result, the scale $t'_*$ under which the logarithmic suppression of $T_c$ becomes visible is usually exponentially low. 

\begin{figure}[!tb]%
\centering
\includegraphics[width=\columnwidth]{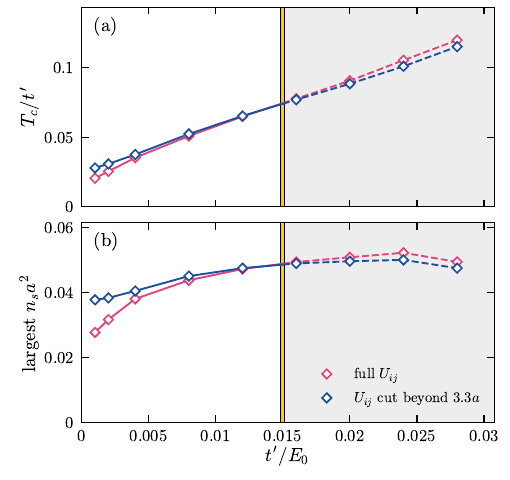}
\caption{Effect of the exponential tail of the interaction between the composite bosons, along the pink line in Fig.~\ref{fig:tc_plane}(b). The pink curves use the full interaction $U_{ij}$. For the blue curves, we drop every $U_{ij}$ below $1\%$ of the largest one, i.e., all the repulsion beyond $3.3a$. (a)~$T_c/t'$, as in Fig.~\ref{fig:tc_plane}(e). (b)~The largest $n_s$, as in Fig.~\ref{fig:tc_plane}(d). In both panels, the curves are dashed and the region is shaded outside the controlled region.}
\label{fig:boson_cut}
\end{figure}

Below we show numerically that the logarithmic suppression is indeed due to the exponential tail, and that it is negligible unless $t'$ is extremely low.
For that, we cut off the tail of $U_{ij}$ and compute $T_c$ along the line $t_h=E_0$ again (see Fig.~\ref{fig:boson_cut}). Namely, we drop every $U_{ij}$ below $1\%$ of the largest one, i.e., all the repulsion beyond $3.3a$. For $t'\gtrsim0.01E_0$, the results with and without the tail agree within $5\%$. At very small $t'$, the largest $n_s$ keeps falling with the full $U_{ij}$ (pink curve), whereas it levels off with the tail cut off (blue curve). Therefore, this result confirms our expectation: the logarithmic suppression is due to the tail in $U_{ij}$ and it is a small correction that occurs only below about $t'_*$.

\clearpage%
\bibliographystyle{apsrev4-2}
\bibliography{ref}%

\end{document}